\documentclass[aps,pre,twocolumn,groupedaddress,showpacs,amsfonts,10pt,%
tightenlines,floatfix]{revtex4-1}

\usepackage[utf8]{inputenc}
\usepackage{graphicx}
\usepackage{amsmath,amssymb}
\usepackage{upgreek}

\let \varpsi \psi
\let \phi \varphi
\let \vec \mathbf
\let \pi \uppi
 \usepackage{xcolor}

\begin{document}
\title{Shapes of sedimenting soft elastic capsules 
in a viscous fluid}

\author{Horst-Holger Boltz}
\email[]{horst-holger.boltz@udo.edu}

\author{Jan Kierfeld}
\email[]{jan.kierfeld@tu-dortmund.de}

\affiliation{Physics Department, TU Dortmund University, 
44221 Dortmund, Germany}

\date{\today}

\begin{abstract}
Soft elastic capsules which are driven through a
viscous fluid  undergo shape deformation coupled to their motion. 
We introduce an iterative solution scheme which  
couples hydrodynamic boundary integral methods and 
elastic shape equations to find the
stationary axisymmetric shape and the velocity
of an  elastic capsule moving  in  a viscous fluid at low 
Reynolds numbers.  We  use this  approach 
to systematically study dynamical shape transitions of 
capsules with Hookean stretching and bending energies
and  spherical rest shape 
sedimenting under the influence of gravity or centrifugal forces. 
We find three types of possible axisymmetric stationary shapes for 
sedimenting capsules with fixed volume: 
a pseudospherical state, a pear-shaped state, and  buckled shapes.
Capsule shapes are controlled by two dimensionless
parameters, the F{\"o}ppl-von-K\'{a}rm\'{a}n number
characterizing the  elastic properties and a Bond number
 characterizing the driving force.
For increasing gravitational force the spherical 
shape transforms into a pear shape. 
For very large bending rigidity (very 
small F{\"o}ppl-von-K\'{a}rm\'{a}n number) 
this transition is discontinuous  with  shape hysteresis.  
The corresponding transition line terminates, however,  in a critical 
point, such that the discontinuous  transition is not present at  
typical  F{\"o}ppl-von-K\'{a}rm\'{a}n numbers of synthetic capsules.
In an additional bifurcation, buckled shapes occur 
upon increasing the gravitational force. 
This type of instability should be observable for generic
synthetic capsules.
All shape bifurcations can be resolved in the 
 force-velocity relation of sedimenting capsules, where 
 up to three  capsule shapes with different velocities 
 can occur for the same  driving  force. 
 All three types of possible axisymmetric stationary shapes 
 are stable with respect to rotation during sedimentation. 
Additionally, we study capsules pushed or pulled 
by a point force, where we always find capsule 
shapes to transform smoothly 
without bifurcations. 
\end{abstract}


\maketitle

\section{Introduction}

On the microscale, 
the motion and deformation of closed soft elastic objects 
through a viscous fluid at low Reynolds numbers, 
either by a driving force or 
 in a hydrodynamic flow, represents an important 
problem with numerous applications, for example, for 
elastic microcapsules \cite{Barthes-Biesel2011}, 
red blood cells \cite{Fedosov2014,Freund2014}, or vesicles 
moving in capillaries, deforming in shear flow \cite{Abreu2014}, 
or sedimenting under gravity \cite{huang2011}. 
Another related system are droplets moving in a viscous 
fluid \cite{clift1978bubbles,Stone1994}.

The analytical description and the simulation of 
soft elastic objects moving in a fluid are 
challenging problems as the hydrodynamics of the  fluid 
 is coupled to the 
elastic deformation of the capsule or vesicle
\cite{Pozrikidis1992,Pozrikidis2001,Barthes-Biesel2011}. 
It is important to recognize that this coupling is mutual:
On the one hand, hydrodynamic forces deform a soft capsule,  a vesicle,
or a droplet.
On the other hand, the  deformed capsule, vesicle, or droplet changes 
the boundary conditions for the fluid flow. 
As a result of this interplay, the soft object deforms and 
takes on characteristic shapes; eventually there are dynamical transitions
between different shapes as a function of the driving force 
or flow velocity. Such shape changes  have important 
consequences for applications or biological function.

In the following, we investigate the stationary shapes of 
sedimenting elastic capsules, 
which are moving in an otherwise quiescent incompressible 
 fluid because of a homogeneous body force, which can be either 
the gravitational or  centrifugal force.  
We focus on  capsules on the microscale with 
radii $\sim$$10\,{\rm \mu m} ~\text{to}~1\,{\rm mm}$, a simple spherical 
rest shape, and sufficiently 
slow sedimenting velocities, such that the  
Reynolds numbers are small and  we can use Stokes flow. 

We use the problem of sedimenting capsules to 
introduce a new iterative method
to calculate efficiently axisymmetric stationary capsule shapes, which 
iterates between a  boundary integral method to solve the 
Stokes flow problem for given capsule shape 
\cite{Pozrikidis1992,Pozrikidis2001,Barthes-Biesel2011} and capsule
 shape equations to calculate the capsule shape in a  given fluid 
velocity field. The method does not capture the dynamic evolution 
of the capsule shape but converges to its {\em stationary} shape. 
The method  can be easily generalized to other 
types of driving forces apart from 
homogeneous body forces. We show results for a driving point force
but, in principle, the method can also be applied to 
 self-propelled  capsules.
The methods allows us to investigate instabilities and bifurcations 
of the final stationary capsule shape with high accuracy.

Elastic capsules have a closed elastic membrane, which  is a
two-dimensional solid that  can support in-plane  shear stresses  and 
inhomogeneous stretching stresses with respect to its rest 
shape and has a bending rigidity.
The theoretical concept of an elastic capsule is rather general 
such that chemistry and nature provide many examples. 
Synthetic capsules, for example, with polymer gel interfaces, can 
be fabricated by various methods and with tunable mechanical 
properties  and have numerous applications
in encapsulation and delivery \cite{Meier2000,Neubauer2013}.
There are also important biological examples of capsules, 
such as spherical (icosahedral) viruses, red blood cells, or 
artificial cells consisting of a lipid bilayer and a cortex
from filamentous proteins such as actin 
\cite{Pontani2009,Tsai2011,Carvalho2013,Schaefer2013}, 
spectrin \cite{Lopez-Montero2012}, 
amyloid  fibrils \cite{Saha2013}, or MreB filaments \cite{Maeda2012}. 
More generally, the cortex of all 
animal cells consisting of the plasma membrane 
and the underlying actin (and spectrin) filament network can 
be regarded as an elastic capsule. 
These  types of capsules differ, however, in elastic 
properties.

Synthetic capsules from polymer gels 
typically  have a spherical rest shape and sizes 
in the micrometer to millimeter range with a micrometer
thick shell. Their  shell materials  can be described by 
isotropic elasticity of thin shells \cite{Vinogradova2006,Neubauer2013}. 
Viruses are much smaller nanometer-sized objects but can also be 
described by thin shell 
isotropic elasticity \cite{Lidmar2003}. Red blood cells
acquire a nonspherical discoid rest shape 
under physiological conditions. Their elasticity is no longer 
 isotropic but the bilayer membrane dominates
the bending rigidity, whereas the spectrin skeleton contributes 
a relatively small shear elasticity, while the 
bilayer membrane remains practically unstretchable
\cite{Mukhopadhyay2002,LimHW2002}. 
For artificial or biological cells with an actin cortex the 
elastic properties are governed by the filamentous network
and strongly vary with the mesh size: At large mesh size,
the bending modulus is relatively large, 
 resembling red blood cell elasticity, at small
mesh size the bending modulus becomes small compared 
to the stretching modulus  resembling the situation 
for a thin  isotropic elastic shell.

Sedimentation has already been studied  for vesicles, both 
numerically \cite{Boedec2011a,Boedec2012,Boedec2013,ReySuarez2013}
 and experimentally \cite{huang2011}.
As opposed to elastic capsules, vesicles are bounded 
by a lipid bilayer, which is a  two-dimensional fluid surface.
 For sedimenting vesicles, pearlike and egglike 
 shapes  have been  observed experimentally  \cite{huang2011}. 
 Numerically, depending on the 
 initial configuration, pear shapes, 
  banana shapes,
 or parachute shapes are found \cite{Boedec2011a,Boedec2012}.
The banana shape exhibits  circulating surface flows
\cite{Boedec2011a,Boedec2012}.
 At sufficiently high velocities, 
 tethering instabilities occur both in experiments \cite{huang2011} and 
 simulations \cite{Boedec2013}.

Sedimentation has also been studied for red blood cells, 
which constitute a special 
type of soft elastic capsule, which is unstretchable and has 
 a nonspherical  biconcave rest shape.
For sedimenting red blood cells shape transitions also have been found.
 Early experiments on centrifuged red blood cells 
  \cite{Corry1978,Corry1978a}
 show shapes developing tails during centrifugation. 
 In Ref.\ \citenum{Hoffman2006}, a sequence of shapes
 from  the biconcave rest shape to cup-shaped and 
 bag-shaped cells has been observed as a function of sedimenting time. 
 Extensive multiparticle collision dynamics (MPCD)
  simulations on sedimenting red blood cell models \cite{Peltomaki2013}
 produced  a dynamic shape diagram, which exhibits 
  tear drop shapes, parachute (or cup-shaped) blood cells, 
 and fin-tailed shapes.

Shape transitions of sedimenting spherical elastic 
capsules have not yet been studied although
the spherical rest shape is  prepared more 
easily in applications involving synthetic microcapsules.
Our iterative method allows us to completely characterize 
capsule shapes 
as a function of their elastic properties and the 
driving force. 
The shapes depend on two dimensionless parameters:
(i)  the  F{\"o}ppl-von-K\'{a}rm\'{a}n number describes the 
typical ratio of stretching to bending energy and characterizes 
the elastic properties of the capsule and (ii) 
the Bond number describes the strength of the driving force 
relative to elastic deformation forces. 
We map out all stationary sedimenting shapes in a shape 
dynamic diagram  parameter 
plane spanned by  F{\"o}ppl-von-K\'{a}rm\'{a}n and Bond number. 
We can use this diagram to 
identify the accessible sedimenting shapes for different types of 
capsule elasticity, such as isotropic shells, red blood cell
elasticity, or semiflexible polymer network  elasticity.

Another important property of soft objects driven 
through a liquid 
is the  relation between the 
driving force  and the resulting velocity of the 
  elastic objects. 
For strictly spherical sedimenting capsules 
this relation is the simple Stokes' law.
Shape transformations of a deformable 
sedimenting object should reflect in qualitative changes or even bifurcations 
in the force-velocity curves.  
This important aspect has  only been  poorly studied for 
droplets, vesicles, and red blood cells. 
Some results have been obtained for quasispherical vesicles in 
Ref.\ \citenum{Boedec2012}.
  MPCD simulations for red blood cells 
 showed no specific signs of shape transitions 
  in the force-velocity  relations \cite{Peltomaki2013}, experimental 
 measurements for vesicles \cite{huang2011} are also not 
 precise enough to find such features.
We will calculate  the force-velocity relation for sedimenting 
elastic capsules with high accuracy in this paper, such that 
we can address this issue and identify 
shape bifurcations of  sedimenting spherical capsules by 
their force-velocity relation.

\section{Methods}

There are many simulation methods which have been applied to the 
dynamics of  vesicles or capsules in viscous fluids, 
such as  particle-based hydrodynamic simulation techniques 
such as MPCD \cite{Noguchi2005},
 dissipative particle dynamics \cite{Fedosov2011}, 
and Lattice Boltzmann simulations \cite{Dupin2007}
or  boundary integral methods \cite{Barthes-Biesel2011}.
These techniques  simulate the  full time-dependent dynamics 
using a triangulated representation of vesicles or capsules;
stationary states are obtained in the  long-time limit. 
Here we introduce an  iterative boundary integral method coupled 
to shape equations for axisymmetric shapes, which 
directly converges to  stationary shapes  without 
simulation of the real dynamics.
Instead, we exploit axisymmetry to avoid triangulated representations 
and get an  efficient iterative 
method based on boundary integrals for the fluid and shape equations  
for the capsule.

We limit ourselves to axisymmetric capsule shapes resulting also 
in axisymmetric viscous fluid flows. Then
the elastic surface of the sedimenting capsule 
has to be at rest in its stationary shape
without any surface flows. Therefore, also the fluid inside the 
capsule will be at rest (in the frame moving with the capsule). 
We will obtain the stationary capsule shape by solving a shape equation 
in curvilinear cylindrical coordinates and solve for the Stokes flow
of the surrounding fluid by applying a boundary integral method.
In an iterative procedure we converge 
to the stationary shape, where the fluid forces onto the capsule and 
the boundary conditions for the fluid flow posed by the capsule shape 
are consistently  incorporated. 
We perform this iterative procedure for a given  driving gravitational 
(or centrifugal) force  and obtain the corresponding 
capsule 
sedimenting velocity from the condition of force balance with the 
total fluid force onto the capsule surface.

\subsection{Geometry}

\begin{figure}[tbh]
\centering
 \includegraphics{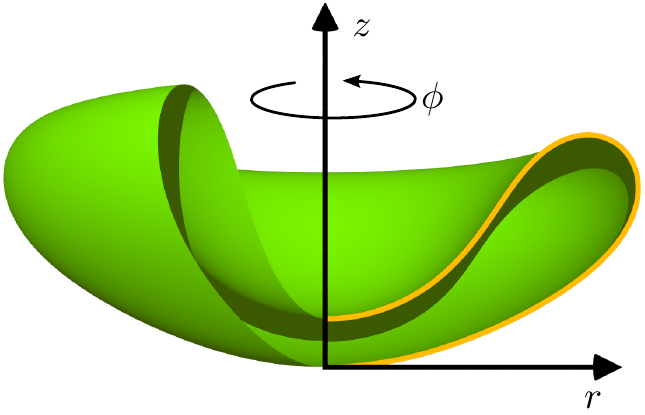}
 \caption{ (color online)
  Example of an axisymmetric shell with coordinates $z$, $r$, and $\phi$.
The shell is generated by revolution of the generatrix
(orange thin line) that is calculated by  the shape equations.
}
 \label{fig:schnitt}
\end{figure}

We work in cylindrical
coordinates to directly exploit the axisymmetry. 
The axis of symmetry is called
$z$, the
distance to this axis $r$, and the polar angle $\phi$, see Fig.\
\ref{fig:schnitt}. The shell is
given by the generatrix $[r(s),z(s)]$ which is parametrized in arc length $s$
(starting at the lower apex with $s=0$ and ending at the upper apex with $s=L$).
The unit tangent  vector $\vec{e}_s$ 
 to the generatrix at $[r(s),z(s)]$ defines
an angle $\psi$ via $\vec{e}_s=(\cos \psi, \sin \psi)$, which can be used to
quantify the orientation of a patch of the capsule relative to the axis of
symmetry.

\subsection{Hydrodynamics}

We want to calculate the flow field of a viscous incompressible fluid 
around an axisymmetric capsule of given fixed shape.
 For the calculation of the flow field the deformability 
of the capsule is not relevant, and the capsule can be viewed 
as a  general  immersed body of revolution $B$. 
For the calculation of the capsule shape
and 
the determination of its 
sedimenting velocity, we  only  need  to  calculate the  
 surface forces onto the capsule which are generated by the fluid flow. 

In the limit of small Reynolds numbers the stress tensor $\sigma$ in a
stationary fluid on which no external body forces are exerted is given by the
stationary Stokes equation \cite{HappelBrenner1983}
\begin{equation}
 \nabla\cdot \sigma = 0 \text{.}
\label{eq:Stokes}
\end{equation}
In Cartesian coordinates, the stress tensor is given by
$\sigma_{ij} = -p \delta_{ij} + 2 \mu e_{ij}$, where 
  $p$ is the hydrodynamic pressure, $\mu$ the viscosity,
 and $e_{ij} = \left(  {\partial_j u_i}+{\partial_i u_j} \right)/2$ is 
 the rate of deformation tensor
for a fluid  velocity field $\vec{u}$.
Additionally, for an incompressible fluid, the continuity equation
$\nabla\cdot \vec{u} = 0$ holds. From the stress tensor one can infer the
surface force density 
$\vec{f}=\sigma \cdot \vec{n}$ onto the capsule 
 by means of the local surface normal $\vec{n}$.

In the rest frame of the sedimenting axisymmetric capsule and 
with   a ``no-slip'' condition at the capsule surface $\partial B$, 
we are looking for axisymmetric solutions that have a given
flow velocity $\vec{u}^\infty$ at infinity and vanishing velocity on the
capsule boundary. 
In the laboratory frame, $-\vec{u}^\infty$ is  the sedimenting 
velocity of the capsule in the stationary state.
Therefore, $\vec{u}^\infty$ has to be determined 
by balancing the total gravitational pulling force 
and the total hydrodynamic drag force on the capsule.

In the laboratory
 frame we are looking for solutions with vanishing pressure and
velocity at infinity. The Green's function for these boundary conditions is
the well-known Stokeslet (also called Oseen-Burgers tensor), that is, the fluid
velocity $\vec{u}$ at $\vec{y}$ due to a point-force $-\vec{F}_p$ 
acting on the fluid at $\vec{x}$,
\begin{equation}
 \vec{u}(\vec{y}) = -\frac{1}{8\pi \mu} \mathsf{G}(\vec{y}-\vec{x}) \cdot
\vec{F}_p
\label{eq:uGF}
\end{equation}
with the Stokeslet $\mathsf{G}$ whose elements are in Cartesian coordinates
($x\equiv \lvert \vec{x}\rvert$)
\begin{equation}
 \mathsf{G}_{ij}(\vec{x}) = \frac{\delta_{ij}}{x}+\frac{x_i x_j}{x^3} 
\text{.}
\end{equation}
This is the basis of 
 the boundary integral approach for  a solution of the Stokes
equation, where we want to find the correct distribution of 
point forces to match the ``no-slip'' condition 
on the capsule surface.

 Because of 
 the axisymmetry we can integrate over the polar angle and find
 the velocity due to a ring of point forces with a local 
force density $-\vec{f}$ acting on the fluid,
$\vec{u}(\vec{y}) = -\frac{1}{8\pi \mu} \mathsf{M}(\vec{y},\vec{x}) \cdot
\vec{f}$, 
with a matrix $\mathsf{M}$, which can be calculated by integration of the
Stokeslet $\mathsf{G}$ with respect to the polar angle.
Switching from  forces $-\vec{f}$ acting  on the fluid to forces 
$\vec{f}$ acting on the capsule and 
integrating over all point forces on the surface, 
we find the general solution of the Stokes equation 
for an axisymmetric  point force distribution on the 
surface of the body of revolution $B$,
\begin{equation}
u_\alpha (\vec y) = - \frac{1}{8\pi\mu} \int_C
\mathrm{d}s(\vec{x}) \mathsf{M}_{\alpha\beta}(\vec{y},\vec{x}) f_\beta(\vec{x})
\text{.}
\label{eq:sol1}
\end{equation} 
Here Greek indices denote the component in cylindrical coordinates, i.e.,
$\alpha,\beta= r, z$ ($u_\varphi=f_\varphi=0$ for symmetry reasons).
For these coordinates, 
the elements of the matrix kernel  $\mathsf{M}$  are given
in Appendix \ref{sec:appM}
according to Ref.\ \citenum{Pozrikidis1992}.  
The integration in (\ref{eq:sol1}) runs along the path $C$ given by the 
generatrix, i.e., the cross section of the 
 boundary $\partial B$, with arc length $s(\vec{x})$.
This representation of a Stokes flow in terms of a 
{\em single-layer potential} 
(using only the Stokeslet and not the stresslet) is possible as long
as there is no net flow through the surface of the capsule
\begin{equation}
 \int \mathrm{d}A\, (\vec{u}-\vec{u}_0)\cdot \vec{n} = 0 \text{.}
\end{equation}

According to the ``no-slip'' condition we have (working in the 
laboratory frame)
$\vec{u} = -\vec{u}^\infty$ at every point $\vec y \in \partial B$ on the
surface. This results in the equation
\begin{equation}
 u_\alpha^\infty  = \frac{1}{8\pi\mu} \int_C
\mathrm{d}s(\vec{x}) \mathsf{M}_{\alpha\beta}(\vec{y},\vec{x})
f_\beta(\vec{x})
~~(\mbox{for}~\vec y \in \partial B),
\label{eq:sol2}
\end{equation}
which can be used to determine the 
 surface  force distribution.

To numerically solve the integral for the surface force
$f(\vec{x}_i)$ at a given set of points $\{\vec{x}_i\}$ ($i=1,\ldots,N$) we
employ a simple collocation method, i.e., we choose a discretized
 representation of the function $f_\beta(\vec{x})$ 
and approximate the integral  in (\ref{eq:sol2}) 
by the rectangle method.
This leads to a system of linear equations
\begin{equation}
 \vec{U} = \tilde{\mathsf{M}} \vec{F}
\end{equation}
with the ``super-vectors''
\begin{eqnarray}
 \vec{U} &=& ( u^\infty(\vec{y}_1)_r, u^\infty(\vec{y}_1)_z, \ldots,
u^\infty(\vec{y}_N)_r, u^\infty(\vec{y}_N)_z) 
     \nonumber\\
   &=& (0 , u^\infty, \ldots, 0, u^\infty)    \\
 \vec{F} &=& ( f(\vec{x}_1)_r, f(\vec{x}_1)_z, \ldots,
f(\vec{x}_N)_r, f(\vec{x}_N)_z)
\end{eqnarray}
and a matrix (using numbered indices 
$r\widehat{=}0$, $z\widehat{=}1$ for a more compact notation)
\begin{equation}
 \tilde{\mathsf{M}}_{2i+\alpha-1,2j+\beta-1} = \frac{1}{8\pi\mu}
(s(\vec{x}_{j+1})-s(\vec{x}_{j})) \mathsf{M}_{\alpha\beta}(\vec{y}_i,\vec{x}_j).
\end{equation}
Due to singularities in the diagonal components of $\mathsf{M}$,
 the two sets of
points $\{\vec{x}_i\}, \{\vec{y}_i\}$ must be distinct (or $\mathsf{M}$
otherwise regularized), so  $2N$ points on the surface are needed, and
the solution is the surface force at points $\{\vec{y}_i\}$.  We validate
our regularization in Appendix \ref{app:perrin}.
For use in the shape
equations, we then decompose this surface force into one component 
acting normal to
the surface, the hydrodynamic pressure $p_n$, and another component acting
tangential to the surface, the shear pressure $p_s$
(see Fig.\ \ref{fig:forces}).

\begin{figure*}[tbh]
\centering
 \includegraphics[width=\linewidth]{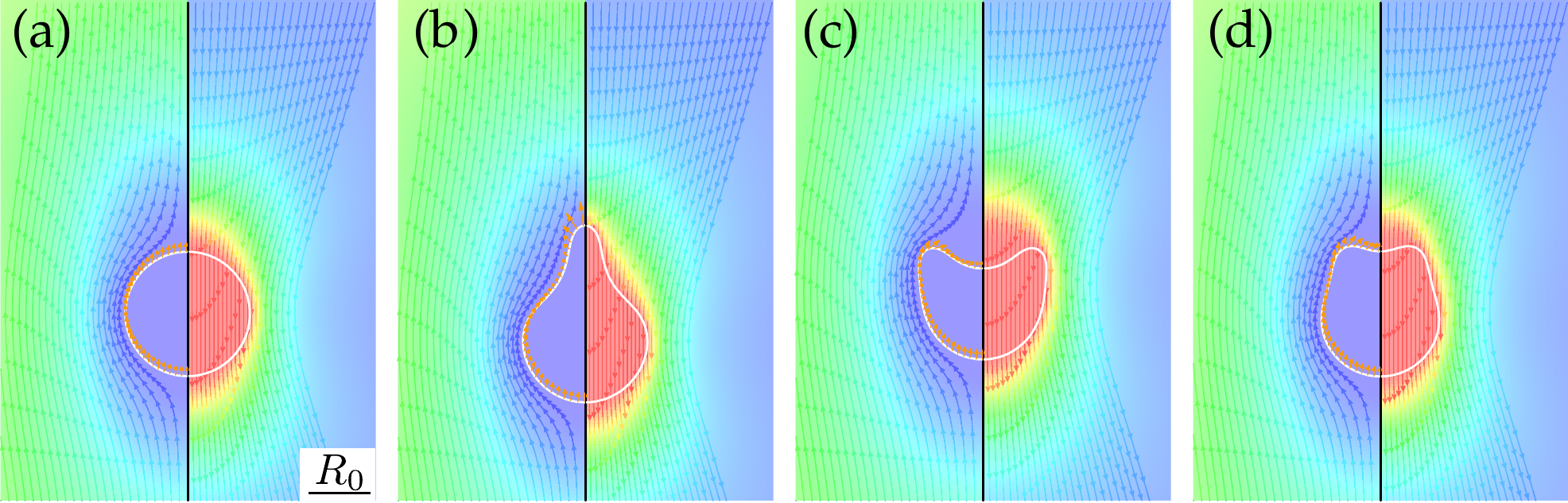}
 \caption{ (Color online)
Fluid velocity field 
   around the four stationary capsule shapes under volume control,
(a) pseudospherical, (b) pear shape, (c) strongly buckled, and 
 (d)   weakly buckled, in the capsule resting frame (left halves)
and in the laboratory frame (right halves).  
 The sedimenting force acts downwards, and 
arrows indicate the direction of flow.
  The arrows are equidistant in time, so  a higher arrow density
   along a line represents a lower velocity. 
  Additionally, the absolute value
   of the velocity is color coded in the background and the hydrodynamic
   surface forces are shown with arrows on the surface. The scale is
     identical  in panels (a)-(d) and given by the  scale bar of length $R_0$
    in (a).
}
 \label{fig:streamlines}
\end{figure*}

We note that we usually restrict our computations
to the bare minimum, i.e., the  surface forces needed for the 
calculation of the capsule shape but, thereby, have  all
necessary information to reconstruct the whole velocity field in the 
surrounding 
liquid as shown in Fig.\ \ref{fig:streamlines}.

Throughout this paper, we use  a ``no-slip'' condition, 
that is, the velocity directly
at the surface of the immersed body vanishes in its resting frame. It
is possible, however, 
 to extend this boundary integral approach to
incorporate a prescribed velocity field on the surface in the resting frame of
the capsule \cite{Pozrikidis1992}.
This will allow us to generalize  the approach
 to model  {\it active} swimmers \cite{Lauga2009,degen2014}
in future work.

\subsection{Equilibrium shape of capsule}

The sedimenting capsule is deformed by the hydrodynamic 
stresses from the surrounding fluid.
We calculate the equilibrium shape of the capsule by a 
set of shape equations, which have been derived in 
Refs.\ \citenum{Knoche2011,Knoche2013}.  We generalize  these
shape equations to 
include the additional fluid stresses on the capsule, which 
we obtain as the surface force density components 
 $f_r$  and $f_z$ from the hydrodynamic flow
as described in the previous section.
 In order to make our approach self-consistent, the 
 capsule shape $B$ from which hydrodynamic surface forces are 
 calculated has to be identical to  the capsule shape that is 
 obtained by integration of the shape equation under the influence
 of these hydrodynamic surface forces. 
 This is achieved by an iterative procedure,
 which will be explained in the following section.

The shape of a thin axisymmetric shell of  thickness $H$
can be derived from 
non-linear shell theory \cite{LibaiSimmonds1998,Pozrikidis2003}. 
A known reference
shape $[r_0(s_0),z_0(s_0)]$ (a subscript zero refers to a quantity of the
reference shape;  $s_0 \in [0,L_0]$ is the arc length 
of the reference shape)
is deformed by hydrodynamic forces exerted by the viscous flow.
Each point $[r_0(s_0), z_0(s_0)]$ is mapped onto a point 
$[r(s_0), z(s_0)]$ in the deformed configuration, 
which induces  meridional and
circumferential stretches, $\lambda_s = ds/ds_0$ and 
$\lambda_\varphi = r/r_0$, respectively. 
The arc length element $ds$ of the deformed
configuration is  $ds^2 = [r'(s_0)^2 + z'(s_0)^2] ds_0^2$.

\begin{figure}[tbh]
\centering
 \includegraphics{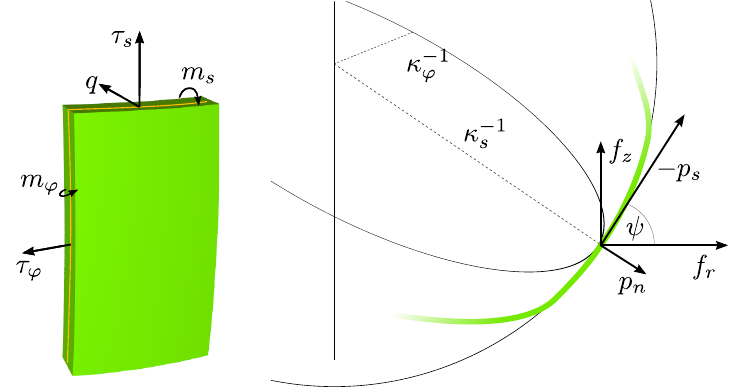}
 \caption{ 
(Color online)
   Left:  Tensions and
   bending moments acting on a shell segment (thickness $H$). 
   Right: Part of the generatrix (green thick line) with the 
  decomposition of external
   surface force densities
  into the coordinate components ($f_r$, $f_z$) as well as 
   into normal
   and tangent components ($p_n$, $p_s$). Also shown are the two principal
   curvature radii $\kappa_s^{-1}$ and $\kappa_\varphi^{-1}$ with
   corresponding  osculating circles.
}
\label{fig:forces}
\end{figure}

The shape of the deformed  axisymmetric shell
 is given by the solution of a system of first-order
differential equations, henceforth referred to as the shape equations. 
Using the notation of Refs.\ \citenum{Knoche2011,Knoche2013} 
these can be written as 
\begin{eqnarray}
 r'(s_0) &=& \lambda_s \cos{\varpsi}~,~~
    z'(s_0) = \lambda_s \sin{\varpsi}~,~~
    \varpsi'(s_0) = \lambda_s \kappa_s \nonumber\\
 \tau'_s(s_0) &=& \lambda_s \left( \frac{\tau_\varphi - \tau_s}{r} \cos\varpsi
+ \kappa_s q + p_s\right) \nonumber\\
 m'_s(s_0) &=& \lambda_s \left( \frac{m_\varphi - m_s}{r} \cos\varpsi
- q + l\right)\nonumber\\
 q'(s_0) &=& \lambda_s \left( -\kappa_s \tau_s - \kappa_\varphi \tau_\varphi -
\frac{q}{r}
\cos\varpsi + p \right) \text{.}
\label{eq:shape}
\end{eqnarray}
The additional 
quantities appearing in these shape equations are shown in 
Fig.\ \ref{fig:forces} and defined as follows:
The angle $\varpsi$ is the slope angle between 
the tangent plane to the deformed
shape and the $r$ axis, $\kappa_\varphi$ is the
circumferential curvature, $\kappa_s$ the meridional curvature.
The meridional and circumferential stresses are denoted by
$\tau_s$ and $\tau_\varphi$, respectively;
$q$ is the transverse shear stress, $p$ the total normal pressure, 
$p_s$ the shear pressure, and $l$ the external stress couple. 

The first three of the shape equations (\ref{eq:shape})
follow from geometry, the last
three equations express force and moment equilibrium.
 For a derivation of these equations
we refer to Refs.\ \citenum{LibaiSimmonds1998,Pozrikidis2003,Knoche2011}.
In order to close the Eqs.\ (\ref{eq:shape}), additional 
geometrical relations 
\begin{equation}
 \kappa_\varphi = \frac{\sin\varpsi}{r}
    ~,~~\kappa_s = \frac{d\psi}{ds}~,~~
 \lambda_\varphi = \frac{r}{r_0}
\label{eq:geom}
\end{equation}
and  constitutive stress-strain relations that depend on the
elastic law governing the shell material are needed. 
The elastic law  will be discussed in the following paragraph.

The  normal pressure  
\begin{equation}
 p=p_0+p_n - g_0 \Delta \rho z 
\label{eq:p}
\end{equation}
(i.e., the pressure difference $p\equiv p_{\rm in}-p_{\rm out}$
between inside and outside pressure),
the shear-pressure $p_s$, and the stress couple $l=p_s H/2$ (the fluid
inside the capsule is at rest) are given externally
by hydrodynamic and external driving forces; $g_0$ is the gravitational
acceleration 
and $\Delta \rho=\rho_{\rm in}-\rho_{\rm out}$ the density difference 
between the fluids inside and outside the capsule. 
Note that we measure the gravitational 
 hydrostatic pressure 
$- g_0 \Delta \rho z$ relative to the lower apex, 
for which we can always choose $z(0)=0$.

The pressures $p_n$ and $p_s$ are the normal and tangential forces 
per area, which are generated by the surrounding fluid.
The surface force density vector $\vec f = f_z \vec e_z + f_r\vec e_r$,
which has been calculated in the previous section,
can also be expressed in terms of $p_n$ and $p_s$, 
 $\vec f = p_n \vec{n} - p_s \vec{e}_s$ ($\vec{n}$ 
and $\vec{e}_s$ are  normal and tangent unit vectors to the generatrix).
Using the decompositions $\vec{n}=-\cos \varpsi \vec{e}_z 
+ \sin \varpsi \vec{e}_r$
and $\vec{e}_s=\sin \varpsi \vec{e}_z +\cos \varpsi \vec{e}_r$ with 
the slope angle $\varpsi$, we find:
\begin{align}
p_n &=  \vec{f}\cdot \vec{n} = f_r\sin\varpsi - f_z\cos\varpsi \\
p_s &= - \vec{f}\cdot \vec{e}_s =-f_r\cos\varpsi -  f_z\sin\varpsi 
\text{.}
\label{eq:pnps}
\end{align}
This is also illustrated in Fig.\ \ref{fig:forces}.

All remaining 
quantities in the shape equations (\ref{eq:shape}) 
follow from constitutive (stress-strain) relations that depend on the
elastic law, which will be discussed in the following paragraph. 
In addition to the elastic law, there might be global
constraints. We consider here only one geometric constraint to the
shape, namely a fixed volume $V=V_0$ 
(due to an incompressible fluid inside the
closed capsule). We introduce
the conjugated Lagrange parameter $p_0$, 
the static pressure difference between
the interior and exterior fluids, which also enters 
the pressure $p$ in Eq.\ (\ref{eq:p}). Additionally, the shape has
to be in global force equilibrium, to which the velocity 
of the capsule is the conjugated parameter.

\subsubsection{Elastic law and reference shapes}

Within this work we use a Hookean energy density with a 
spherical resting shape
to model  deformable capsules.
As a measure for the deformation of the capsule we introduce
the meridional and circumferential strains
\begin{equation}
 e_s = \lambda_s -1~,~~
 e_\varphi = \lambda_\varphi -1 
\end{equation}
as well as the meridional and circumferential bending strains
\begin{equation}
 K_s = \lambda_s \kappa_s -{\kappa_{s}}_0 ~,~~
 K_\varphi = \lambda_\varphi \kappa_\varphi -{\kappa_\varphi}_0 
\text{.}
\end{equation}
The elastic energy we use is Hookean, i.e.,  a quadratic form
in these strains. More precisely, the surface energy density $w$, which measures
the elastic energy of an infinitesimal patch of the (deformed) surface divided
by the area of this patch in the undeformed state, is given by
\begin{eqnarray}
 w_s &=& \frac{1}{2} \frac{Y_{\rm 2D}}{1-\nu^2} 
       \left(e_s^2 + 2\nu e_s e_\varphi + e_\varphi^2\right)\nonumber\\
&~&+\frac{1}{2}E_{B} 
       \left( K_s^2 + 2\nu K_s K_\varphi + K_\varphi^2 \right) 
\label{eq:Hook}
\end{eqnarray}
with the two-dimensional Young modulus $Y_{\rm 2D}$, 
which defines the tension (energy
per surface) scale, the
 bending modulus $E_{B}$
and the (two-dimensional) Poisson ratio
$\nu$ (assuming equal Poisson ratios for bending and stretching). 
The fundamental length scale is
the radius $R_0$ of the spherical rest shape (with 
$R_0={\kappa_{s}}_0^{-1}$  and $L_0 = \pi R_0$).
Tensions and bending moments derive from the surface energy (\ref{eq:Hook}) 
by $\tau_s = \lambda_\varphi^{-1} \partial w/\partial e_s$ and 
$K_s = \lambda_\varphi^{-1} \partial w/\partial K_s$  (and two more 
analogous relations with indices $s$ and $\varphi$ interchanged),
which gives the missing constitutive stress-strain relations
for the shape equations (\ref{eq:shape}). 
For Hookean elasticity and a spherical rest shape, 
the resulting set of shape equations has been  solved for a 
purely  hydrostatic  pressure $p=p_0$ and  $p_s=0$  
in Ref.\ \citenum{Knoche2011}.

The Hookean elastic model for a spherical rest shape 
contains five parameters, $Y_{\rm 2D}$, $R_0$, $E_B$, $\nu$, and  $H$,
characterizing capsule size and elastic properties. 
In Sec.\ \ref{sec:resc} below, we will eliminate 
two parameters, $Y_{\rm 2D}$ and $R_0$, by choosing our scales
of  energy (or tension) and length appropriately.
The two-dimensional Poisson ratio is bounded
by $\nu\in[-1;1]$, and we 
reduce our parameter space by always using $\nu=1/2$. This
way our Hookean energy density has the same behavior for small stresses as
the more complex Mooney-Rivlin functional \cite{LibaiSimmonds1998}. 
The thickness $H$ enters the shape equations
 only via the stress couple $l$ and,
thus, only  weakly influences  the resulting shape. Assuming
that the shell can be treated as a thin shell made from an isotropic elastic
material, the bending modulus is directly related to the shell
thickness $H$ by
\begin{equation}
 E_B = \frac{Y_{\rm 2D} H^2}{12(1-\nu^2)} = \frac{1}{9}Y_{\rm 2D}H^2 \text{.} 
\label{eq:eb}
\end{equation}
Thus, we have one remaining free parameter to change the elastic properties,
the  bending modulus $E_B$.

The Hookean elastic energy law we use here is well suited to describe the
deformation behavior of soft elastic capsules. For other systems different
energies 
might be more adequate (e.g.\
 Helfrich bending energy and a locally inextensible surface for
vesicles). As pointed out above, a different choice of
elastic law enters the formalism through the constitutive relations and, thus,
 does not require changes in our method on the  conceptual level.

\subsubsection{Solution of the shape equations}

The boundary conditions for a shape that is closed and has no kinks at its
poles are
\begin{equation}
  r(0) = r(L_0) = \psi(0) = \pi - \psi(L_0) = 0,
\label{eq:rpsi0}
\end{equation}
and we can always choose $z(0)=0$. 
 For the solution of the shape equations it is important
that there is no net force on the capsule as the shape equations are derived
from force equilibrium.
The condition that  the total hydrodynamic drag force 
equals the total gravitational   force
determines  the resulting sedimenting velocity.
If hydrodynamic drag and
 gravitational pull cancel each other in a stationary state, 
there is no remaining point force at
the poles needed to ensure equilibrium and, thus,
\begin{equation}
 q(0) = q(L_0) = 0 \text{.}
\label{eq:q0}
\end{equation}
The shape equations have (removable) singularities at both poles; therefore, a
numerical solution has to start at both poles requiring $12$ 
boundary conditions ($r,z,\psi, \tau, m, q$ at both poles), from which we know
$7$ [by Eqs.\ (\ref{eq:rpsi0}) and (\ref{eq:q0}) and $z(0)=0$]. 
The $5$ remaining parameters can be determined by a shooting method using
that the solution starting at $s_0=0$ and the one starting at $s_0=L_0$ have to
match continuously  in the middle, 
which gives $6$ matching conditions ($r,z,\psi, \tau, m,
q$). This gives an overdetermined nonlinear set of equations which we solve
iteratively using linearizations. 
However, as in the static
case \cite{Knoche2011}, the existence of a solution to the resulting system of
linear equations (the matching conditions) is ensured by the existence of a
first integral (see Appendix \ref{sec:appfirst}) 
of the shape equation. In principle, this first integral could be
used to cancel out the matching condition for one parameter (e.g., $q$),
and, thus, we 
have $5$ independent equations to determine $5$ parameters, and the system
is not genuinely overdetermined. We found the
approach using an overdetermined system to be better numerically 
tractable, where  we ultimately used a multiple shooting method
including several matching points between the poles.
Throughout this work we used a
fourth-order Runge-Kutta scheme with step width 
  $\Delta s_0=5\times 10^{-5} R_0$.

Using these boundary conditions, it is straightforward to see that the
shape equations  do not allow for a solution whose shape is the reference
shape, unless there are no external loads ($p_s=p=l=0$). There will be
solutions arbitrarily close to a sphere, which we call pseudospherical.

\subsection{Iterative solution of shape and flow
 and determination of sedimenting  velocity}

\begin{figure}[ht]
 \includegraphics{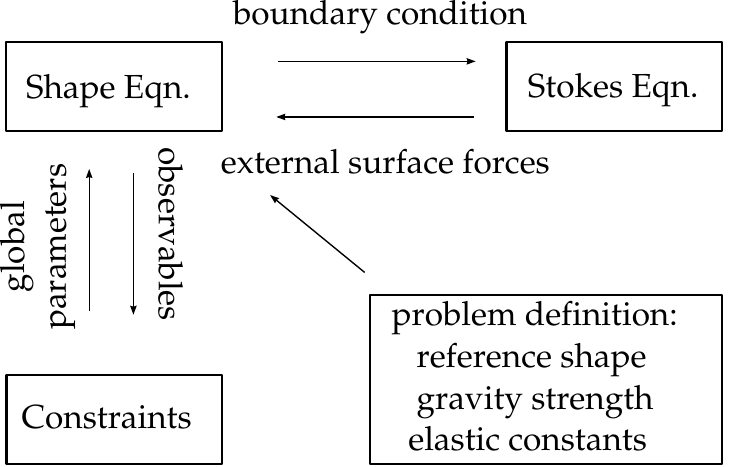}
 \caption{ 
  Iterative scheme for the solution to the problem of
elastic capsules in Stokes flow as explained in the text. 
}
 \label{fig:feedback}
\end{figure}

We find a joint solution to the shape equations and
the Stokes equation by solving them separately and iteratively,
as illustrated in the scheme in  Fig.\ \ref{fig:feedback},
to converge to the desired solution:
We assume a fixed axisymmetric shape and calculate the resulting 
hydrodynamic forces on the capsule  for this shape. 
Then, we use the resulting  hydrodynamic  surface force density 
to calculate a new deformed shape. 
Using this new shape we re-calculate the hydrodynamic surface forces
and so on. 
We iterate until a fixed point is reached. 
At the fixed point, our approach is self-consistent, i.e., the 
capsule shape from which hydrodynamic surface forces are 
calculated is identical to  the capsule shape that is 
obtained by integration of the shape equation under the influence
of exactly these  hydrodynamic surface forces.

For each capsule shape during the iteration, we can determine 
its  sedimenting velocity $u= |\vec{u}^\infty|$ by 
requiring that the total hydrodynamic drag force 
equals the total gravitational force,
\begin{equation}
  g_0 \Delta \rho V_0  = - \int_C\mathrm{d}s(\vec{x}) 2\pi r(\vec{x}) 
              f_z(\vec{x}) 
\label{eq:sedimenting_u}
\end{equation}
The integration runs along the path $C$ given by the 
generatrix, i.e., the cross section of the 
 boundary $\partial B$, with arc length $s(\vec{x})$.
By changing $u=|\vec{u}^\infty|$ we can adjust the hydrodynamic 
drag forces on the right hand side to achieve equality for a 
given capsule shape during the iteration. 
During the iteration $u$ will converge to the proper sedimenting 
velocity for the stationary state.

The Stokes equation is linear in the velocity, and we
can just scale the resulting surface forces accordingly, if we change the
velocity parameter $u^\infty$.  
In this way, the global force balance can be treated the
same way as other possible constraints like a fixed volume. 
Numerically, it is
impossible to ensure the exact equality of the drag and the drive forces 
in Eq.\ (\ref{eq:sedimenting_u}). 
Demanding a very small residual force difference 
 makes it difficult to find an adequate
velocity, and a too-large force difference
 makes it impossible to find a solution with
small errors at the matching points. 
Global force balance  (\ref{eq:sedimenting_u}) 
is equivalent to the condition that the total 
force in axial direction vanishes, 
\begin{equation}
   X(L) = -2\pi \int_0^L \mathrm{d}s\, r (p\cos\psi+p_s\sin\psi) = 0 \text{,}
\label{eq:XL}
\end{equation}
see Eqs.\ (\ref{eq:p}) and  (\ref{eq:pnps}) and Fig.\ \ref{fig:forces}.
Interestingly, the total axial force 
is directly related to the existence of a   first integral  of the shape
equations, which is discussed in the Appendix \ref{sec:appfirst}, 
see Eq.\ (\ref{eq:appX}).
 To monitor global force balance numerically, we use a criterion $\lvert X(L)
 \rvert < 10^{-5}$, which turns out to be a good compromise for a numerical
 force balance criterion.

The iteration starts with a given (arbitrary) stress, e.g.,
 one corresponding to
the flow around the reference shape. 
For the resulting initial 
capsule shape,  the Stokes flow is
computed and the resulting stress is then used to start the iteration. 
If, during the iteration,  the new and old stress
differ strongly it might be difficult to find the new shooting parameters 
for the capsule shape and
the right sedimenting velocity starting at their old values. 
To overcome this technical problem, we use a convex
combination $\sigma = \alpha \sigma_{\text{new}} + (1-\alpha)
\sigma_{\text{old}}$ of the two stresses and slowly increase the fraction
$\alpha$ of
the new stress until it reaches unity. 
 The iteration continues until the change 
 (monitored in pressure and velocity) within one iteration 
 cycle is sufficiently
 small.
In total, this allows for a joint solution of the shape equations and the 
Stokes equation to find the stationary shape and the sedimenting velocity 
at rather small numerical cost.

 If there are multiple stationary solutions at a given gravitational
strength (see discussion of the shape diagram \ref{fig:vol}
in Sec.\ \ref{ref:shapediagram} below), 
 the iterative procedure 
  obviously converges to only  one shape. 
Which stationary shape  is selected depends on the initial configuration.
A stationary shape can be continuated to  different  parameter 
values by slightly changing the control parameters and using the 
former stationary shape as new initial shape.

In order to find all branches of stationary shapes we first 
generate  capsule shapes for a fixed (artificial) flow field for the 
whole range of gravitational fields at a  fixed bending rigidity. 
These serve as initial shapes from which we 
iterate until a stationary shape with the correct flow field
is reached.
At low bending rigidities, this procedure generates 
(all four) different classes of stationary shapes depending on the 
gravitational force.  
We then try to continuate all different classes of stationary shapes 
to the whole range of 
 gravitational forces and, afterwards, to the whole 
range of  bending rigidities.
This allows us to obtain a full set of
 solutions and identify all different branches.


\section{Results}

In this section we present the results for stationary axisymmetric 
sedimenting shapes 
and stationary sedimenting velocities 
as obtained by the fixed point iteration method. 
We
focus on  the sedimentation of capsules under volume control
($V=V_0=\text{const}$) in the main text 
and present additional results for  pressure control ($p_0=\text{const}$)
in Appendix \ref{sec:pressure}.
Additionally, we show the results for a capsule that is driven 
(or pulled) by a localized point 
force rather than a homogeneous body force.

\subsection{Control parameters and non-dimensionalization}
\label{sec:resc} 

In order to identify the relevant control parameters 
 and reduce the parameter space, in the remainder of the paper 
we introduce dimensionless quantities 
by measuring 
lengths in units of the radius $R_0$ of the spherical rest shape of the
capsule, energies in units of $Y_{\rm 2D}R_0^2$ (i.e., tensions
in units of   $Y_{\rm 2D}$),
 and times in units of $R_0 \mu/Y_{\rm 2D}$. 
This results in the following set of
control parameters for the capsule shape:
\begin{enumerate}

\item  
Our elastic law is fully
characterized by the  dimensionless bending modulus 
or its inverse,  the F{\"o}ppl-von-K\'{a}rm\'{a}n
number \cite{LibaiSimmonds1998}, 
\begin{equation}
  \tilde{E}_B\equiv \frac{E_B}{Y_{\rm 2D} R_0^2} =  \frac{1}{\gamma_{\rm FvK}}.
\label{eq:tildeEB}
\end{equation}
In general, elastic properties also depend on the Poisson ratio $\nu$ 
but  we limit ourselves to $\nu=1/2$ as explained above.
 Using the thin-shell result (\ref{eq:eb}) the dimensionless
 bending
modulus $\tilde{E}_B = (H/9R_0)^2$ also determines the shell 
thickness $H$  for an  isotropic shell material.

\item 
 The sedimentation motion of the capsule is determined 
by the strength of the gravitational (or centrifugal) 
pull $g_0 \Delta \rho$
(the gravitational force density), 
which we measure in units of $Y_{\rm 2D}/R_0^2$ in the following.
The dimensionless
gravitational force $g$  then takes the form of a Bond number, 
\begin{equation}
  g =  {\mathrm Bo} \equiv \frac{ g_0 \Delta \rho R_0^2}{Y_{\rm 2D}},
\label{eq:Bo}
\end{equation}
where the characteristic 
elastic tension $Y_{\rm 2D}$ is used instead of a liquid surface tension. 

\item 
The static pressure $p_0$ within the capsule
is  measured in units of $Y_{\rm 2D}/R_0$ in the following.

\item
The resulting
sedimenting velocity $u$  is measured in units of $Y_{\rm 2D}/\mu$.
\end{enumerate}
We conclude that the resulting stationary sedimenting 
shapes of the capsule are fully  determined by 
 two dimensionless control parameters if the capsule 
volume is fixed:
(i)  dimensionless bending modulus $\tilde{E}_B$ or its inverse, 
 the  F{\"o}ppl-von-K\'{a}rm\'{a}n number $\gamma_{\rm FvK}= 1/\tilde{E}_B$ 
  characterizes the elastic properties of the capsule
(by the typical ratio of  bending to stretching energy)  and (ii) 
the Bond number describes the strength of the driving force 
(relative to elastic deformation forces).
If the capsule pressure is controlled rather than its volume, 
the dimensionless pressure $p_0$ 
provides a third parameter, see Appendix \ref{sec:pressure}.
Because  stationary shapes are independent of time, they 
do not depend on the solvent viscosity.

\subsection{Shape diagram}
\label{ref:shapediagram}

\begin{figure*}[tbh]
 \includegraphics[width=0.95\linewidth]{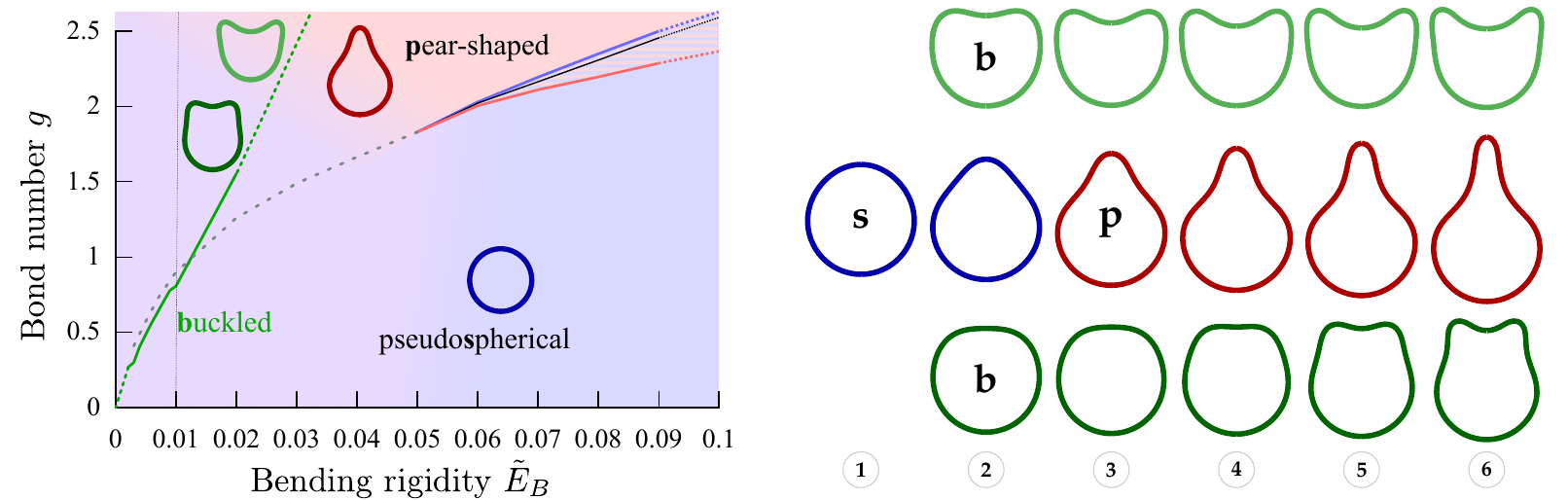}
 \caption{ (Color online)
Left: Stationary axisymmetric shapes for a sedimenting 
Hookean capsule with spherical
reference shape in the parameter plane of the two control parameters, 
 the dimensionless bending modulus $\tilde{E}_B$ (the
inverse of the F{\"o}ppl-von-K\'{a}rm\'{a}n number $\gamma_{\rm FvK}$) 
and the dimensionless 
gravity $g$ (the Bond number $\mathrm{Bo}$).
We find pseudospherical (blue), pear-shaped (red), and a pair of 
strongly (light green) and weakly (dark green) buckled shapes.  
The black line is a line of discontinuous shape transitions
between pear and pseudospherical shape which terminates in 
a critical point beyond which a sphere transforms smoothly into 
the pear shape. 
The blue and red lines are spinodals indicating the limits of stability 
of the pseudospherical and pear-shaped shapes, respectively. 
Above the solid green line  the pair of buckled shapes occur 
in a bifurcation;
the dotted continuation of the solid line signals numerical 
difficulties in following this bifurcation line. 
To which solution the iteration procedure 
converges in  regions with shape coexistence is controlled by 
the initial conditions.
Right: Contours of all four types of shapes
for $\tilde{E}_B=0.01$, i.e., along the vertical black dotted line,
in the range $g =  0.7~\text{to}~1.7$. 
The gravity strength $g$ increases
from left to right in steps of $\Delta g=0.2$.
}
 \label{fig:vol} 
\end{figure*}

The spherical rest shape has a dimensionless volume $V= 4\pi/3$. 
For sedimentation under volume control, we use this as the 
fixed volume $V_0 = 4\pi/3$.
 In principle, it is also possible to prescribe volumes that differ
 from the rest shape volume. This can  be done using 
 pressure control, for which we present 
  results  in Appendix  \ref{sec:pressure}.
 For the static case, the only stationary shape at fixed volume 
 is the strictly spherical  shape, regardless of the 
 bending modulus. This differs for a sedimenting capsule, 
 which displays various shape transitions already at fixed volume.

We summarize our findings regarding the axisymmetric 
sedimenting shapes in the 
shape diagram Fig.\ \ref{fig:vol} in the $(\tilde{E}_B, g)$ plane 
of the  two control parameters.
We investigated stationary axisymmetric capsule shapes 
for gravity strengths (Bond numbers) 
up to $g=g_{\text{max}}=2.5$
(usually using steps of $\Delta g=0.005$) and 
bending moduli $\tilde{E}_B = 10^{-3}~\text{to}~10^{-1}$ corresponding to
F{\"o}ppl-von-K\'{a}rm\'{a}n numbers $\gamma_{\rm FvK} = 10~\text{to}~1000$. 
For high Bond and F{\"o}ppl-von-K\'{a}rm\'{a}n numbers, the
iteration procedure becomes numerically more demanding (higher forces
ask for finer discretization of the shape equations, 
 smaller distances between
solutions in parameter space ask for a more thorough fixed point search), but
remains possible in principle.
We will discuss further below in detail, which parameter regimes in the 
$(\tilde{E}_B, g)$ plane are accessible for 
different types of capsules and sedimenting driving forces.

Overall, we identify three classes
of axisymmetric shapes, which 
are  shown in Fig.\ \ref{fig:streamlines}:
\begin{enumerate}
\item 
  Pseudospherical shapes, which remain convex and close to the rest shape.
  For pseudospherical shapes the velocity  is still given  by 
  the result for a perfect sphere, which is 
   Stokes' law  
  $u\approx  F/6\pi \approx 2g/9$
   (in dimensionless form), to a good approximation. 
  Likewise, the 
   pressure due to gravity is  $p_0 \approx g\langle z \rangle\approx g$.
\item 
  Pear shapes, which have a convex upper apex but 
  develop  an axisymmetric indentation at the side of the capsule.
  For high gravitational driving force (high sedimentation velocities)
  the pear shape resembles a tether with a high curvature at
  the upper apex.
\item 
  Buckled shapes, which  develop a concave  axisymmetric  indentation
  at the upper apex of the capsule. 
  For fixed capsule volume, 
  these buckled shapes always occur in pairs of a weakly buckled 
  shape  with a shallow and narrow indentation and a strongly buckled shape
  with a deep and wide indentation at the upper apex. 
\end{enumerate}

Our solution method  allows us to  
 identify all  bifurcations or transitions between these 
classes of shapes as shown in Fig.\ \ref{fig:vol}. 
Upon increasing the gravity $g$ or decreasing the 
 bending rigidity $\tilde{E}_B$, 
the  stationary sedimenting pseudospherical shape transforms  
 into a pear shape.
The black line in  Fig.\ \ref{fig:vol}
is a line of {\it discontinuous} shape transitions
between pseudospherical and pear shape, which terminates in 
a critical point at 
\begin{equation}
  g_c\simeq 1.85 ~\mbox{and}~  \tilde{E}_{B,c}\simeq 0.05.
\label{eq:crit}
\end{equation}
At the discontinuous shape transition we find  hysteresis with 
two spinodal lines  (red and blue lines) indicating the limits of stability 
of pseudospherical and pear-shape shapes. 
For smaller  bending rigidities $\tilde{E}_{B}<\tilde{E}_{B,c}$,
 the  spherical shape  transforms {\it smoothly} into  the pear shape
upon increasing gravity.
We locate the transition lines between the sphere and pear shape 
by the condition of equal sedimenting velocity (for the discontinuous
transition)  or maximal sedimenting velocity (for the smooth crossover)
as explained in the next section.
Discontinuous buckling transitions are also known from static 
spherical shells \cite{LL7, pogorelov1988}, for example,
as a function of the external static pressure or an osmotic pressure
\cite{Knoche2014}. 
That a line of discontinuous buckling transitions 
terminates in a critical point as a function of the elastic properties 
of the capsules is, however, 
an unknown phenomenon for  static buckling transitions.

Within the smooth crossover regime between sphere and pear shapes
at small bending rigidities $\tilde{E}_{B}<\tilde{E}_{B,c}$, the 
pair of stationary 
buckled shapes occurs in a bifurcation upon increasing 
the  gravity 
strengths $g$ above  a critical  threshold given by the  solid 
green line in Fig.\ \ref{fig:vol}. 
Also the spherical or pear shape  remains a stable solution 
above the green line, such that we have three possible stationary 
sedimenting shapes in this parameter regime. 
Which of these shapes is dynamically selected in an 
actual experiment, 
depends on the initial conditions. 
The dotted continuation of the solid green line  signals numerical 
difficulties in following this bifurcation line.

The iterative method allows us to determine stationary sedimenting 
shapes in the shape diagram  Fig.\ \ref{fig:vol} 
very efficiently. In order to continuate the branch of pseudospherical and
pear-shaped solutions for a fixed bending rigidity $\tilde{E}_B=0.01$ in Bond
number steps of 
$\Delta g = 0.005$ in the  range $g=0,...,2$, we need a
computing time of approximately 6 h on a four core Intel Xeon 
processor (3.7 GHz).

\subsection{Force-velocity relation}

\begin{figure}[htb]
 \includegraphics{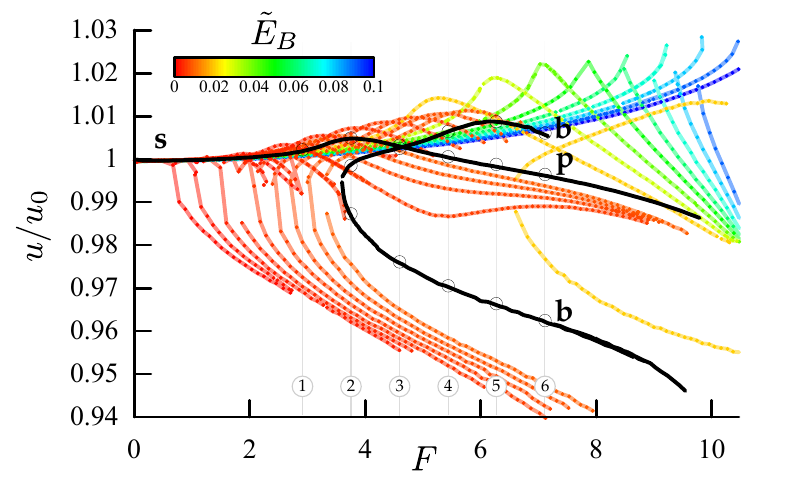}
 \caption{ (Color online)
Force-velocity relations for  sedimenting capsules under volume
control. The plot shows
the velocity relative to the (dimensionless) 
  velocity 
$u_0 = 2g/9$  of a perfect  sphere
with the same volume (Stokes' law) 
as a function of the (dimensionless) total external force $F=4\pi g/3$.
The
bending modulus $\tilde{E}_B$ is color coded. 
Two distinct branches can be distinguished: 
 The solution branches corresponding to pseudospherical and 
pear shapes start  at zero force. 
A pair of  solution branches for strongly or weakly buckled shapes
 starts at a nonzero threshold force. 
The branches for $\tilde{E}_B = 0.01$ are marked in black
as a typical example. The capsule shapes 1--6 in  Fig.\  \ref{fig:vol}
   are realized  at the points marked by circles.
It is also apparent that the transition
 from pseudospherical to pear shapes  is
continuous for $\tilde{E}_B \lessapprox 0.05$ and becomes discontinuous 
for $\tilde{E}_B \gtrapprox 0.05$.
}
 \label{fig:sed}
\end{figure}

All dynamic shape transitions reflect 
in  bifurcations  in the force-velocity relations.
In Fig.\ \ref{fig:sed}, we show the force-velocity relations between 
sedimenting velocity $u$ and gravity $g$ for all 
stationary axisymmetric shapes and for a color-coded 
range of bending rigidities
$\tilde{E}_B =  10^{-3} ~\text{to}~10^{-1}$.
The total driving force $F$ can be computed using the total drag from the
fluid corresponding to the 
 right hand side of Eq.\  (\ref{eq:sedimenting_u})
or as the product of the total mass difference 
 and the sedimenting acceleration
$F=g_0 \Delta \rho V_0$  or $F=4\pi g/3$ (in dimensionless form)
corresponding to the left-hand  side of Eq.\ (\ref{eq:sedimenting_u}).
The leading contribution of the force-velocity relation is 
given by the force-velocity relation of a perfect sphere, 
which is Stokes' law  $u =u_0 = F/6\pi = 2g/9$ (in dimensionless form). 
To eliminate this leading contribution,
   we show  the relative velocity $u/u_0$  in Fig.\ \ref{fig:sed}.
By definition,  the relative velocity is also proportional
 to the {\it sedimentation coefficient}, which 
is defined as $v/g_0$ and a standard quantity in centrifugation and 
 sedimentation experiments \cite{Planken2010}.

 Figure \ref{fig:sed} clearly shows two qualitatively 
different types of force-velocity 
curves.  The force-velocity curves corresponding to  pseudospherical or 
pear shapes  with positive curvature at both apexes 
start at zero force, whereas the solution curves corresponding to the 
weakly and strongly buckled shapes bifurcate 
 at a nonzero threshold force 
with a vertical tangent and have two branches:
The weakly buckled shapes with a narrow indentation 
have a higher sedimenting velocity and 
correspond to the upper branch, and the strongly  buckled shapes have a lower
sedimenting velocity and correspond to the lower branch. 

We note that, because of the $g$ dependence of $u_0$, the absolute 
sedimenting velocities $u$ are increasing with increasing gravity
for both buckled shapes,
although $u/u_0$ is decreasing for the strongly buckled shapes. 
Above the threshold force for the buckled shapes, three different 
axisymmetric sedimenting  capsule shapes can occur 
with different sedimenting velocities 
for  the same gravitational driving  force. 
Which of these shapes is dynamically selected in an 
actual experiment,
depends on the initial conditions.

\begin{figure}[htb]
 \includegraphics[width=0.95\linewidth]{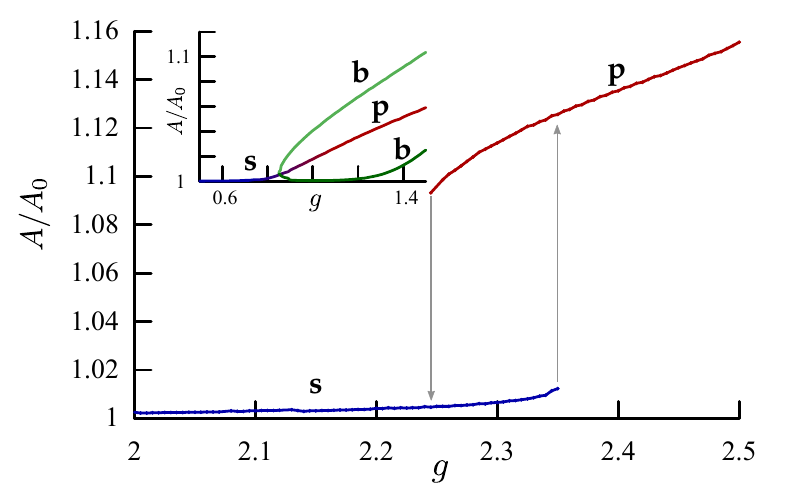}
 \caption{ (Color online)
Reduced capsule area  as a function of the dimensionless 
gravity strength $g$ for a  high
 bending modulus $\tilde{E}_B=0.08$. There is an interval where
pseudospherical and pear-shaped solutions with different area 
coexist with hysteretic switching
upon increasing and decreasing $g$. 
Inset:  For  $\tilde{E}_B=0.01$, the area changes continuously  
from pseudospherical to pear shapes. For the pair of weakly and strongly 
buckled solutions the area bifurcates at a critical $g$. 
}
 \label{fig:fbif}
\end{figure}

The force-velocity curves for the   pseudospherical or 
pear shapes  allow us to detect how the  transition 
between spherical and pear shapes  evolves into a discontinuous 
transition: Figure \ref{fig:sed} shows that  these force-velocity curves 
 develop a {\em cusp} close to the critical 
point for  $\tilde{E}_B \approx E_{B,c} \simeq 0.05$, whereas they 
remain smooth for $\tilde{E}_B < E_{B,c}$.
For $\tilde{E}_B > E_{B,c} $, we find two overlapping and 
intersecting  velocity branches, which corresponds to the 
characteristic  hysteretic velocity switching
in  a discontinuous transition. 
In fact, we identify the discontinuous 
transition line in Fig.\ \ref{fig:vol}  by these velocities:
  If the two branches
 coexist, then we localize the transition at the crossing of the two
  $u(g)$ curves.
For $\tilde{E}_B < E_{B,c}$, the velocity curve still exhibits a
maximum, which we can use to define the crossover line 
between the pseudospherical and pear shape as it is shown in 
the shape diagram Fig.\ \ref{fig:vol}. 
The discontinuous nature of the sphere-pear transition 
is confirmed  by monitoring other quantities, 
such as the  capsule area, as a 
 function of the gravity $g$. As illustrated in  Fig.\ \ref{fig:fbif},
the area clearly exhibits hysteretic switching at the 
transition. 
The pair of weakly and strongly buckled solutions appears 
in a bifurcation above a critical driving force. Also this 
bifurcation reflects in a corresponding bifurcation 
of the capsule area as shown in the inset of  Fig.\ \ref{fig:fbif}.

\subsection{Transition mechanism}

Qualitatively, the shape transformations from a spherical shape 
into a pear shape or into buckled shapes
 are a result of buckling instabilities of a hydrodynamically
stretched capsule at the upper part of the capsule. 
Upon increasing the 
gravitational force $g$ the capsule acquires a higher sedimenting velocity
and stretches along the $z$ direction. 
This stretching is essential as it provides the excess area 
necessary for a shape transformation from a spherical rest shape,
which has the minimal area for the fixed volume.

The numerical results show that 
the fluid flow generates a negative (compressive) contribution  $p_n$
to the interior pressure at the bottom part of the capsule and a 
positive (dilatational) 
 contribution at the upper part for all types of stationary shapes
(see the orange arrows in Fig.\ \ref{fig:streamlines}). 
The  gravitational hydrostatic pressure contribution 
$- g_0 \Delta \rho z$, on the other hand,
generates a negative contribution on the upper part as compared 
to a vanishing contribution on the lower apex (by choice
of the $z$ coordinate). Together with the positive volume pressure 
$p_0$, this results in a compressive negative
 total normal pressure at the upper part and 
dilatational positive total pressure at the lower part. 
Both for pear and buckled 
shapes, the negative total normal pressure at the upper part 
of the capsule, which is mainly 
caused by the gravitational hydrostatic pressure
contribution,  is the reason to develop indentations. 
In the transition from pseudospherical to pear-shaped  capsules
an axisymmetric indentation develops at the side of the capsule (see 
the shapes in the middle row in Fig.\ \ref{fig:vol}),
because 
for an elongated stretched capsule, this part of the capsule has 
lower curvature and, thus, less stability with respect to buckling. 
At higher driving force $g$ and sedimenting velocities, the indentation 
can also form at the upper apex  (see 
top and bottom rows in Fig.\ \ref{fig:vol}), and the pair of 
buckled configurations become stable solutions.

An approximative limit for the stability of the  spherical shape can be 
given in terms of the classical 
 buckling pressure \cite{LL7,LibaiSimmonds1998, pogorelov1988}
$p_c = -4\sqrt{\tilde{E}_B}$ in dimensionless units.
We first note that the height of a sphere is
$z_{\text{max}}\approx 2$ and that the static pressure contribution needed to
maintain a constant volume  against the hydrostatic 
 pressure is
 $p_0 \approx g \overline{z}$, where $\overline{z}\approx 1$ is the 
$z$-component of the center of mass. This gives an
effective compressive pressure $p_0-g z_{\text{max}} \approx -g$ 
at the upper apex
(the compressive hydrodynamic contribution $p_n>0$ is smaller 
and can be neglected compared to relative to 
  hydrostatic and static pressure).
If this  compressive pressure exceeds the 
classical  buckling pressure, i.e.,
$g \gtrsim |p_c|=  4\sqrt{\tilde{E}_B}$, then the spherical shape is  unstable
with respect to indentations  at the upper side of the capsule,
which leads to the pear shape.  
A parameter dependence
\begin{equation} 
  g \propto \sqrt{\tilde{E}_B}
\label{eq:spherepear}
\end{equation}
describes well the boundary between spherical and pear shape in 
the shape diagram Fig.\ \ref{fig:vol}.

The termination of the line of  discontinuous transitions 
between sphere and pear shapes terminating in a critical 
point is a result of the deformation of the capsule 
by the fluid flow prior to the shape transition:
Increasing the gravity $g$ gives rise to a hydrodynamic stretching of the 
upper part of the capsule. The smaller the dimensionless bending rigidity 
$\tilde{E}_B$, the smaller is the meridional curvature $\kappa_s$ in 
the upper part. For soft capsules, the meridional curvature 
vanishes {\em before} the effective compressive pressure becomes 
comparable to the buckling pressure. 
(see, for example, the second shape in the middle row
 in Fig.\ \ref{fig:vol} for $\tilde{E}_B=0.01$).
For a flat shell segment, however, buckling becomes continuous 
and does not require a threshold normal pressure. 
Rigid capsules, on the other hand,  remain curved upon increasing 
the gravity $g$  up to the
discontinuous buckling induced by the normal pressure in the 
upper part of the capsule.

For the  bifurcation line of the two buckled shapes in 
the shape diagram Fig.\ \ref{fig:vol}, 
we find an approximately linear dependence 
$g \propto \sqrt{\tilde{E}_B}^{1.7}$
from fitting our numerical results.
Currently, we  have no simple explanation for this result.

\subsection{Rotational stability}

We always assumed axisymmetric shapes that perform a rotation-free
sedimentation, i.e., stability with respect to rotation or tilt of the shape. 
 The question of whether the sedimentation motion of an
axisymmetric rigid body is rotationally stable can be reduced
\cite{HappelBrenner1983} to the question whether the so-called center of
hydrodynamic stress lies above or below the center of mass; both points have
to be on the symmetry axis. 
In general, sedimenting bodies will tilt in a way
that aligns the connection vector of the centers of mass and hydrodynamic
stress with the direction of gravity \cite{TenHagen2014} or start to
rotate. The center of hydrodynamic stress is the point about which the
translational and rotational motions decouple. 

\begin{figure}
 \centering
 \includegraphics[scale=1.]{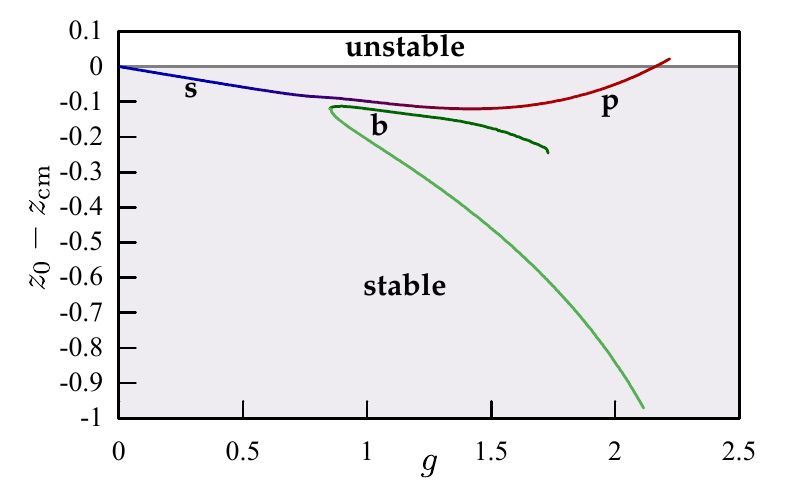}
 \caption{ (Color online)
  Linear analysis concerning rotational
   stability for $\tilde{E_B}=0.01$. 
  We show the (dimensionless) difference $z_0-z_{\text{c.m.}}$ between 
   the center of hydrodynamic stress $z_0$ and 
    the center of gravity $z_{\text{c.m.}}$ as a function of 
  the dimensionless gravity strength $g$ for 
   all three solution  branches 
  [(blue) pseudospherical, (red) pear-shaped, (green) buckled]. 
  For $z_0-z_{\text{c.m.}}<0$, i.e., below the gray line, 
    the shape is linearly stable against out-of-axis rotations.
  Only  solutions with a very pronounced tether extrusion at high $g$ 
  are unstable.
 }
\label{fig:rotstab}
\end{figure}

In Appendix \ref{sec:approt}, we present a stability argument 
that only uses the information available to us. 
The result of this analysis is exemplarily shown in Fig.\ \ref{fig:rotstab}
for $\tilde{E_B}=0.01$. 
We find that the center of hydrodynamic stress $z_0$ lies almost
always below the center of gravity $z_{\text{c.m.}}$ for all
three classes of shapes, pseudospherical, pear-shaped, and buckled
such that the shapes are linearly stable against out-of-axis rotations.
Only  solutions
in the pear-shape class for very high gravity forces $g\gtrsim 2.2$ 
 are unstable. These shapes exhibit  a very pronounced tether-like
 extrusion.

\subsection{Accessible parameter space}

With the shape diagram in Fig.\ \ref{fig:vol} we have a complete 
overview of theoretically possible shapes and transitions between 
them in the plane of the two control parameters,
  the dimensionless bending rigidity $\tilde{E}_B$ or its inverse, 
  the F{\"o}ppl-von-K\'{a}rm\'{a}n number $\gamma_{\rm FvK}=1/\tilde{E}_B$,
  and the Bond number ${\rm Bo} = g$. 
There are three limitations, however, to the accessible parameter 
range.

First,  our analysis is limited to the regime of low 
Reynolds numbers because we used the Stokes equation to describe 
  the fluid flow. The Reynolds number for the sedimenting capsule
is given by $\mathrm{Re}= v\rho R_0 /\mu$.
  To leading order the sedimenting velocity is given by Stokes' law
 $v = g_0 \Delta\rho V/6\pi \mu R_0$  resulting in 
  $\mathrm{Re}= 2g_0\Delta\rho \rho R_0^3/9\mu^2$.
A  criterion $\mathrm{Re}<1$ limits the accessible Bond numbers 
to 
$g<g_{\text{max, Re}} \sim 9 \mu^2 /2\rho R_0 Y_{\rm 2D}$.
For   water as solvent we have 
$g_{\text{max, Re}} \sim 5\times
 10^{-9} R_0^{-1} Y_{\rm 2D}^{-1}\, {\rm N}$. 
This  can be 
easily increased by a factor of 100 in a 
more  viscous solvent.
For a capsule of size $R_0 \sim 10 {\rm \mu m}$ in water, the criterion 
 $g<g_{\text{max, Re}}$
corresponds to a condition that the acceleration should remain  smaller 
than 5000 times the standard gravity.

Second, 
the sedimenting force is very small for gravitation;
it is much larger for a centrifuge but also modern ultracentrifuges 
are limited to accelerations about $10^6$ times the standard gravity
\cite{Planken2010}. 
With a typical density difference of 
$\Delta \rho = 10^2 \,\mathrm{kg}/\mathrm{m^3}$ 
($10\%$ of the density of water) 
we find that the Bond number is limited in available 
centrifuges to 
$g<g_{\text{max, cf}}\sim 10^9 R_0^2 Y_{\rm 2D}^{-1} \,\mathrm{N}/\mathrm{m}^3$. 

Both of these constraints give an upper bound for the 
 accessible Bond number range. 
The constraints are size dependent, however:
For small capsule sizes $R_0\lesssim  1 \, \mathrm{\mu m}$, 
for example, for viruses,  the 
centrifuge  constraint is more restrictive, whereas
for larger 
capsule sizes $R_0\gtrsim  1 \, \mathrm{\mu m}$, 
the low-Reynolds-number constraint is more restrictive.
The latter case is  typical  for all synthetic capsules, 
red blood cells, or artificial cells consisting of a lipid bilayer 
and a filamentous cortex, which all have capsule 
sizes in the range  $R_0\gtrsim  10~\text{to}~1000 \mathrm{\mu m}$.

Third, it is not possible to realize 
arbitrarily large 
dimensionless bending rigidities $\tilde{E}_B =  E_B/Y_{\rm 2D} R_0^2$ 
or small  F{\"o}ppl-von-K\'{a}rm\'{a}n numbers 
in experimentally available capsules. 
For isotropic elastic shell materials the relative shell thickness
determines this parameter,  $\tilde{E}_B\sim (H/R_0)^2$, 
see Eq.\ (\ref{eq:eb}), and 
values $\tilde{E}_B \gtrsim 0.01$ are  difficult to realize
with synthetic capsules. 
Viruses also follow this law (with a shell thickness 
$H\sim 2 {\rm nm}$), resulting in 
   F{\"o}ppl-von-K\'{a}rm\'{a}n numbers 
$\gamma_{\rm FvK}\gtrsim 60$ or $\tilde{E}_B \lesssim 0.02$
with $\tilde{E}_B$ decreasing with increasing  virus size \cite{Lidmar2003};
small viruses with $R_0 \sim 15 {\rm nm}$ realize the largest values 
of $\tilde{E}_B$.

Red blood cells also have similar values of $\tilde{E}_B$
although they are elastically strongly anisotropic 
because they consist of a lipid bilayer and a spectrin cytoskeleton. 
The liquid  lipid bilayer dominates the 
 bending modulus $E_B \sim 50 k_BT$, whereas
the shear modulus is determined by the filamentous spectrin 
cortex or skeleton.
Filamentous networks typically have low shear moduli, for example  
$G_{\rm 2D} \sim 10^{-6} {\rm N/m}$ for the spectrin skeleton 
of a red blood cell \cite{Lenormand2001}. 
Using $Y_{\rm 2D} \sim 4G_{\rm 2D}$ for the red blood cell,
because the lipid bilayer has a much higher area 
expansion modules 
$K_{\rm 2D} \sim 0.3  {\rm N/m}$, we obtain $\tilde{E}_B \sim 0.005$  
for a  size $R_0 \sim 3 {\rm \mu m}$.
Larger values of $\tilde{E}_B$ could be realized in 
 artificial cells consisting of a lipid bilayer and a cortex
from filamentous proteins for low filament densities in the cortex. 
These  estimates of $\tilde{E}_B$ for red blood cells or 
 artificial cells might, however,  be
misleading because the relevant elastic modulus 
in the elastic 
energy  (\ref{eq:Hook}) is actually $Y_{\rm 2D}/(1-\nu^2)$ rather than $Y_{\rm
  2D}$, which is used in the standard definition of the
F{\"o}ppl-von-K\'{a}rm\'{a}n number and  $\tilde{E}_B$, see Eq.\
(\ref{eq:tildeEB}).
Because of their lipid bilayer,  red blood cells or artificial cells are 
in the unstretchable 
limit $K_{\rm 2D} \gg G_{\rm 2D}$, where $\nu$ approaches unity, 
and  we obtain  $Y_{\rm 2D}/(1-\nu^2)\approx K_{\rm 2D}$. 
Using this modulus in a definition of $\tilde{E}_B$ 
(i.e., in the non-dimensionalization)  leads to 
much smaller values $\tilde{E}_B \sim 10^{-9}$.

The discontinuous sphere-pear transition happens beyond 
a critical point, $\tilde{E}_B > \tilde{E}_{B,c}\simeq 0.05$, see 
Fig.\ \ref{fig:vol}.
Therefore, this peculiar transition is not accessible 
 for typical synthetic capsules.  To observe this transition 
an elastically very anisotropic capsule shell 
material with high bending and 
low stretching moduli would be  needed. 
The transition into buckled shapes, however, should be 
well accessible with generic soft synthetic capsules.

The softness of the shell material is also important 
if high Bond numbers ${\rm Bo} \sim 1$  
are to be realized in the shape diagram Fig.\ \ref{fig:vol}. 
In the presence of the above low-Reynolds-number constraint,
very soft 
shell materials with a small two-dimensional 
Young modulus $Y_{\rm 2D}\lesssim 5\times 10^{-4} {\rm N/m}$ for a 
capsule of size $R_0 \sim 10 {\rm \mu m}$ are needed to reach 
such Bond numbers in water. Typical soft 
synthetic capsules such as OTS capsules have 
much larger moduli $Y_{\rm 2D} \sim 0.1 {\rm N/m}$ \cite{Knoche2013}. 
Red blood cells or artificial cells  with  shear moduli
governed by a soft filamentous network have small moduli 
$Y_{\rm 2D} \sim 10^{-6} {\rm N/m}$, which will allow us to 
reach high Bond numbers.  Again, these estimates 
 might be misleading because the relevant elastic modulus 
in the elastic 
energy is actually $Y_{\rm 2D}/(1-\nu^2)$ rather than $Y_{\rm
  2D}$. If this  modulus is used in the nondimensionalization
and the definition of the Bond number in Eq.\ (\ref{eq:Bo}), 
  Bond numbers become much smaller
because red blood cells or artificial cells are hardly stretchable.

\section{Localized driving forces}

There are other possible external driving forces and self-propulsion
mechanisms \cite{ebbens2010, degen2014}. 
One  extreme case is a very localized external force. 
We study this case by employing a
driving pressure that acts on a small patch on one apex (we used the
criterion $s_0 \leq 0.1$ or $L-s_0 \leq 0.1$, respectively; 
even more localized
pressures  would have to be higher and, thus, require smaller
integration steps). 
For a pushing force, i.e., a force that acts on the stern of the
capsule, this gives rise 
 to strongly indented shapes. They have a 
higher drag than a spherical shape such that the driving force 
has to be higher than in Stokes' law $u\approx  F/6\pi $
   (in dimensionless form) 
 resulting in $F_{\text{drive}} > 6 \pi  u $, whereas for
pulling forces (acting on the bow) the drag is lower as compared to 
 a sphere. 

\begin{figure}[tb]
 \includegraphics{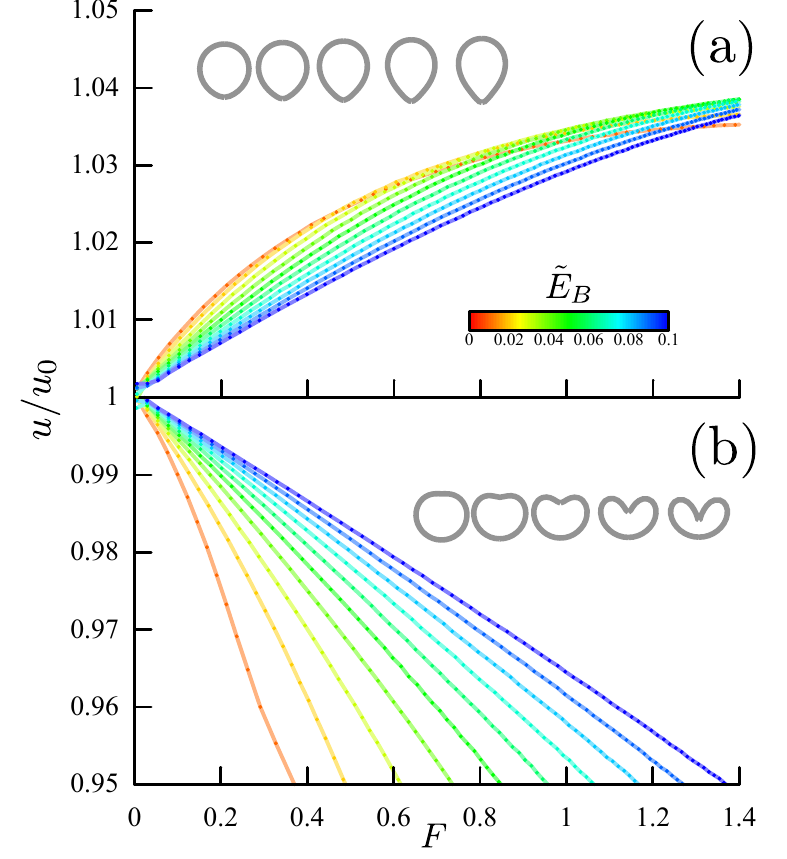}
 \caption{ (Color online)
Force-velocity relations 
for a capsule with fixed volume  driven by 
(a) a localized {\em pulling} force and 
(b)  a localized {\em pushing} force. 
The plots show
the velocity relative to the (dimensionless) 
velocity $u_0=F/6\pi$ of a hard sphere
with same volume as a function of the (dimensionless) total external force. The
insets show
exemplary capsule shapes  for $E_B=0.01$.
}
 \label{fig:pf}
\end{figure}

We show force-velocity relations with some exemplary shapes for these two types
of external 
driving forces  in Fig.\ \ref{fig:pf}. 
As can be seen clearly seen in the force-velocity relations, 
there are {\em no} shape bifurcations for elastic capsules 
driven by localized forces.

One way to apply a localized force experimentally is to
attach beads to the capsule that can be manipulated via optical tweezers,
as it has been done 
(for pulling and pushing types of forces) with giant vesicles in
Ref.\ \citenum{dasgupta2014}.

\section{Discussion and Conclusion}

We introduced a new iterative solution scheme to find 
stationary axisymmetric shapes and velocity
of a deformable  elastic capsule driven through a
viscous fluid by an external force at low Reynolds numbers. 
We focused on homogeneous body forces, i.e., 
sedimenting capsules in gravitation or in a centrifuge,  but also 
demonstrated that other force distributions, such as localized forces, 
can be studied.

The iterative method is sufficiently accurate and fast to 
identify all branches of different shapes and to resolve 
dynamic shape transitions. 
We find a rich dynamic bifurcation behavior for sedimenting elastic 
capsules, even at fixed volume, see Fig.\ \ref{fig:vol} with 
 three types of possible axisymmetric stationary shapes: 
a pseudospherical state, a pear-shaped state, and  buckled shapes.
The capsule can undergo shape transformations  
as a function of two dimensionless parameters. 
The  elastic properties  are characterized by 
a dimensionless bending rigidity $\tilde{E}_B$
[the inverse F{\"o}ppl-von-K\'{a}rm\'{a}n number, see Eq.\
(\ref{eq:tildeEB})], and 
the driving force is characterized by 
the  dimensionless  Bond number [see Eq.\ (\ref{eq:Bo})].

The transition between pseudospherical and pear shape is 
a  discontinuous transition 
with shape coexistence and 
hysteresis for large bending rigidity 
or large Bond number. 
The corresponding transition line terminates, however,  in a critical 
point if the bending rigidity is lowered, a phenomenon which is 
unknown from discontinuous static buckling transitions.
Parameter estimates show that this transition  cannot be 
observed  with standard synthetic capsules because it requires a 
rather large bending rigidity  or low F{\"o}ppl-von-K\'{a}rm\'{a}n number,
which cannot be realized for thin ($H/R_0 < 0.1$) 
shells of an isotropic elastic material. 
Observation of this discontinuous transition requires a material,
which combines high bending rigidity while it  remains
highly  stretchable.

We find an additional  bifurcation into a pair of 
buckled shapes at small bending rigidities 
 upon increasing the gravitational force. 
This bifurcation should be experimentally accessible with synthetic 
capsules in a centrifuge. 
The bifurcations are caused by hydrodynamic stretching, which increases
a compressive 
hydrostatic pressure at the upper apex and  leads to buckling-like 
transitions.

All shape bifurcations can be resolved in the 
 force-velocity relation of sedimenting capsules, where 
 up to three  capsule shapes with different velocities 
can occur for the same gravitational driving  force, see Fig.\ \ref{fig:sed}. 
In an experiment, these shapes are selected depending on the initial 
conditions. 
We also confirmed the stability of all axisymmetric  shapes  with respect 
to rotation around an axis perpendicular to the axis of symmetry, see Fig.\
\ref{fig:rotstab}. 
We find more stationary buckled shapes if we consider sedimenting capsules
where we vary the volume by changing the pressure as it is explained
in more detail in Appendix \ref{sec:pressure}. 
These additional shapes are higher order buckled shapes 
with  more indentations.

It is instructive to discuss  our  results in  comparison 
with sedimenting vesicles \cite{Boedec2011a,Boedec2012,Boedec2013} 
and sedimenting red blood cells \cite{Peltomaki2013}.
Vesicles are bounded by a two-dimensional fluid lipid bilayer 
rather than a two-dimensional solid membrane or shell and 
are virtually unstretchable, i.e., they 
  have a fixed area. Therefore, shape transitions from a spherical shape 
with the minimal area for given volume 
require excess area. For the sedimenting capsules 
this additional area is generated by hydrodynamic stretching. 
For vesicles, excess area can be ``hidden'' in  thermal fluctuations. 
A spherical vesicle without such excess area cannot undergo any 
shape transitions during sedimentation. 
 The pear shape for elastic capsules at high Bond numbers 
is similar to the ``tether'' formation by sedimenting vesicles
\cite{Boedec2013}. However, there are two important differences.
First, the  stretching energy of capsules penalizes
 deformations from the resting shape (whereas vesicles have a liquid surface
 that allows any shape with the correct area), which leads to less extreme and
 pronounced tethers. Moreover, cylindrical tethers are an actual equilibrium
 shape of a membrane under tension, which can coexist 
 with a spherical vesicle  \cite{Lipowsky2005}. This is not the case 
 for an elastic capsule. 
 Secondly, we do not find a bulge or ``droplet'' forming
 at the upper end, by which we mean an increase in the width of the extrusion
 near the end of the tether. This is rooted in the different bending energies:
  Using a Helfrich bending energy that is quadratic in the mean curvature $w_B
 \sim (\kappa_s + \kappa_{\phi})^2$, the negative meridional curvature
 ($\kappa_s<0$) needed to form a drop at the end of the tether is energetically
 favorable, whereas it is not with the Hookean bending energy (\ref{eq:Hook}) 
 we employ.  
  Moreover, elastic capsules cannot develop shapes with circulating 
 surface flows because of their solid membrane. 
Therefore, there is no analog of the banana shape with 
surface flows that has been found in Refs.\ \citenum{Boedec2011a,Boedec2012}
  for vesicles. 

Also sedimenting red blood cells exhibits different shapes;
in MPCD simulations transitions 
between  tear drop shapes, parachute (or cup-shaped) blood cells 
 and fin-tailed shapes \cite{Peltomaki2013} have been found. 
Red blood cells have a non-spherical discocyte rest shape, 
which already provides some excess area as compared to the minimal 
area of a sphere. Therefore, stretching is not essential 
for dynamic shape transitions. 
The initial discocyte shape of red blood cells also gives rise to 
a strong tendency to tilt, which is absent for our initially spherical 
capsules. The parachute shape also tilts by almost $90^{\circ}$ during 
transformation into a tear drop \cite{Peltomaki2013}.
Interestingly, all transitions 
observed in Ref.\ \citenum{Peltomaki2013} did not show any signatures 
in the force-velocity relation or the bending or shear energies. 
For spherical capsules we find clear signatures of all transitions 
in the force-velocity relation in Fig.\ \ref{fig:sed}.
It remains to be clarified in future work whether the MPCD technique is not 
accurate enough or exhibits too many fluctuations to resolve 
these signatures or whether the dynamic transitions of sedimenting 
red blood  cells qualitatively differ, for example, because
tilt plays such a prominent role.

Finally, we studied capsules driven by localized forces, such as a 
pushing or pulling point-like force. For such localized 
driving forces, we always find shapes to transform smoothly 
without any shape bifurcations, see Fig.\ \ref{fig:pf}.

We plan to extend our iterative 
  method to actively swimming capsules in future work \cite{degen2014}. 
One possible swimming mechanism, for example in squirmer 
type swimmer models \cite{blake1971},  is 
a finite velocity field at the capsule  surface 
in its resting frame. Our
solution technique using single-layer potentials remains still applicable as
long as there is no net velocity normal to the surface (which 
indicates
a fluid source or sink within the capsule and is, thus, not relevant).

\section*{Acknowledgments}

We acknowledge financial support by the Deutsche Forschungsgemeinschaft 
via SPP 1726 ``Microswimmers'' (KI 662/7-1).

\appendix
\section{Single-layer potential solution of the 
Stokes equation in  an axisymmetric domain}
\label{sec:appM}

To make the paper self-contained, we give the analytical expressions for the
elements of the matrix kernel 
$\mathsf{M}$ that relates the forces and the velocities
on the surface of the capsule via
\begin{equation}
 u_\alpha^\infty = -\frac{1}{8\pi\mu} \int_C
\mathrm{d}s(\vec{x}) \mathsf{M}_{\alpha\beta}(\vec{y},\vec{x}) 
 f_\beta(\vec{x})~~(\mbox{for}~\vec y \in \partial B),
\end{equation}
see Eq.\ (\ref{eq:sol2}).
The derivation of this matrix kernel 
is briefly outlined in the main text and can
also be found in Ref.\ \citenum{Pozrikidis1992}.

As we are working in cylindrical coordinates $(r,z,\varphi)$ and the problem
is axisymmetric, the matrix has four elements ($\alpha,\beta=r,z$) which can be
explicitly written as 
\begin{align}
\mathsf{M}_{zz}(\vec y, \vec x) &= 2 k \sqrt{\frac{r_x}{r_y}}
    \left(F+\hat{z}^2 E\right)\\
\mathsf{M}_{zr}(\vec y,\vec x) &= 
     \frac{k\hat{z}}{\sqrt{r_xr_y}} \left(F-
   (r_y^2-r_x^2+\hat{z}^2) E \right)\\
\mathsf{M}_{rz}(\vec y, \vec x) &= 
  -k\frac{\hat{z}}{r_y}\sqrt{\frac{r_x}{r_y}}
   \left( F + (r_y^2-r_x^2-\hat{z}^2)  E  \right)\\
\mathsf{M}_{rr}(\vec y, \vec x) &= 
   \frac{k}{r_x r_y} \sqrt{\frac{r_x}{r_y}}\left[
      (r_y^2+r_x^2+2\hat{z}^2) F\right. \nonumber\\
   &\left.-(2\hat{z}^4+3\hat{z}^2(r_y^2+r_y^2)+(r_y^2-r_x^2)^2)E\right]
\end{align}
using the abbreviations
\begin{align}
\hat{z}&\equiv z_x-z_y \\
k^2 &\equiv  \frac{4r_x r_y}{\hat{z}^2+(r_x+r_y)^2} \\
E&\equiv \frac{ E(k)}{ \hat{z}^2+(r_x-r_y)^2} \\
F&\equiv  K(k) 
\end{align}
and the complete elliptic integrals of the first and second kinds
\cite{Abramowitz1972},
\begin{align}
K(k) &= \int_0^{\pi/2} \frac{\mathrm{d}\!x}{\sqrt{1-k^2 \cos^2x}} \\
E(k) &= \int_0^{\pi/2} \mathrm{d}\!x\,\sqrt{1-k^2\cos^2 x}
\end{align}
For the numerical 
computation of $\mathsf{M}$ elliptic integrals of the first
and second kind we use  polynomial approximants 
as given in Ref.\ \citenum{Abramowitz1972}.

\section{Validation of numerical method}
\label{app:perrin}

\begin{figure}[htb]
\includegraphics[width=0.95\linewidth]{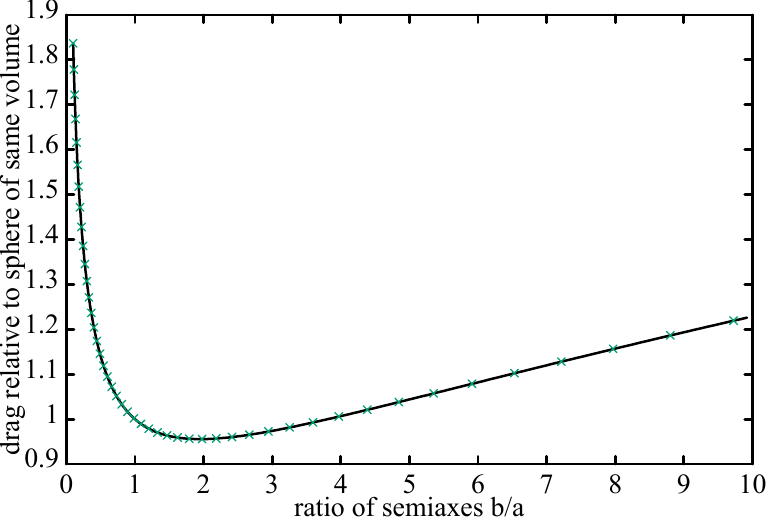}
\caption{  (Color online)
Perrin factor $\Xi$, i.e.,  drag force 
 onto a spheroid in a viscous fluid relative to that onto a
sphere of same volume moving with the same velocity as a function 
 of the ratio of the spheroid semiaxes $\xi=b/a$.  The
theoretical result (\ref{eq:Perrin}) 
due to Perrin is shown as a solid (black) line and our
numerical data as (green) crosses. 
The relative difference is less than $1/1000$
for all values, which validates our numerical approach.
}
 \label{fig:perrin}
\end{figure}

We validate our algorithm and, in particular, our 
numerical treatment of  singularities occurring in the boundary integral 
approach  in  Eq.\ (\ref{eq:sol2}) 
 by comparison 
 with  exact analytical results  for
 the Perrin factors for the total   drag of
a spheroid with semiaxes $a$ in radial and $b$ in  axial direction 
 \cite{HappelBrenner1983,Perrin}. 
A spheroid is an  ellipsoid with two degenerate semiaxes
$a$ in radial direction, 
i.e., it  can be parametrized by
\begin{align}
\vec{r} &= \left(\begin{array}{c} 
a\cos\theta \sin\phi \\ a\sin\theta\sin\phi \\
b\cos\theta \end{array} \right)
\end{align}
with the polar angle $0\leq\phi\leq 2\pi$
and the parametric latitude $0\leq\theta\leq \pi$.

The Perrin factor $\Xi$ is  the
ratio of the total drag force onto a spheroid 
 with semiaxes ratio $\xi = b/a$ 
(for translation along the axial $a$ direction) 
and the drag of a sphere of the same volume
moving with the same velocity.  $\Xi$ is 
known analytically 
 (from calculating  the stream function in ellipsoidal coordinates) 
\cite{Perrin},
\begin{align}
\Xi &=\begin{cases}
   \frac{\xi^{-\frac{1}{3}}}{
\frac{3}{4}\sqrt{\xi_p^2-1}(-\xi_p+(\xi_p^2+1)\operatorname{artanh}(\xi_p^
{-1}))} & \mbox{for}~ \xi\geq 1 \\
\frac{\xi^{-\frac{1}{3}}}{\frac{3}{4}\sqrt{\xi_o^2+1}
(\xi_o-(\xi_o^2-1)\operatorname{atan}(\xi_o^{-1 })} &
\mbox{for}~ \xi<1\end{cases}
\label{eq:Perrin}
\end{align}
with
\begin{align}
\xi_p &\equiv \sqrt{1-\xi^{-2}}^{-1}\\
\xi_o &\equiv \sqrt{\xi^{-2}-1}^{-1},
\end{align}
 for both prolate ($\xi \geq 1$) and oblate ($\xi \leq 1$) spheroids.
We compare our numerical results to this analytical expression 
in Fig.\ \ref{fig:perrin} and find excellent agreement.

\section{First Integral of the shape equations}
\label{sec:appfirst}

Inspired from the known first integral in the static
problem (cf.\ Eqs.\ (17) and  (22) in Ref.\ \citenum{Knoche2011}) we make the
following ansatz for a first integral of the shape equations:
\begin{align}
 U(s) &=  2\pi r\left(q \cos\psi  + \tau_s \sin\psi \right)
   + X =\text{const.}
\end{align}
We are also assuming that the pressure $p$ and the shear pressure $p_s$ can be
written as functions solely dependent on the arc length $s$. The calculation
is rather straightforward: We differentiate using the 
shape equations (\ref{eq:shape}) and the geometric relations 
(\ref{eq:geom}) and get
\begin{align}
 0 &= U'(s) \nonumber\\
 &= 2\pi q\cos^2\psi  - 2\pi r \kappa_s q \sin\psi  \nonumber\\
  &~+ 2\pi r\cos\psi
  \left((-\kappa_s \tau_s - \kappa_\varphi \tau_\varphi -
   \cos\psi \frac{q}{r} +p\right) 
  \nonumber\\
 &~+ 2\pi \tau_s \cos\psi \sin\psi  + 2\pi r
\kappa_s\tau_s\cos\psi
  \nonumber\\
 &~+2\pi r\sin\psi \left(\cos\psi \frac{\tau_\varphi-\tau_s}{r} +
  \kappa_s q + p_s\right)  + X' 
 \nonumber\\
&= 2 \pi r p \cos \psi + 2\pi r p_s \sin\psi + X' = 0
\end{align}
In the second-to-last step most terms cancel each other out. 
Thus, we find 
an ordinary differential equation for $X$, which we can integrate directly:
\begin{align}
 X' &= -2\pi r (p\cos\psi + p_s\sin\psi)\nonumber\\
 X  &= -2\pi \int_0^s \mathrm{d}x\, r(p\cos\psi+p_s\sin\psi)
\label{eq:appX}
\end{align}
Inspecting the behavior at $s=0$  we then deduce $U(s)=U(0) =0$ or 
\begin{align}
 0 &= 2\pi r q \cos\psi  + 2\pi r\tau_s \sin\psi   +X.
\label{eq:U=0}
\end{align}

The first integral $U(s)=0$
 is indeed a generalization of the static quantity, as we see by setting
$p_s=0$, $p=\text{const}$, which gives 
\begin{align}
 X &= -2\pi p \int_0^s \mathrm{d}x\, r\cos\psi = -\pi p r^2,
\end{align}
and, thus, Eq.\ (\ref{eq:U=0}) becomes equivalent to 
Eq.\ (17) in  Ref.\ \citenum{Knoche2011}.

The first integral is related to the 
global axial force balance of the capsule, which can be seen 
by considering $s=L$, where the  $U(s)=U(L)=0$ 
gives $X(L)=0$.
The quantity $X$ contains the contribution to the net force in the $z$
direction
and thus a shape with the desired features (namely $q=0$ at the apexes) must
have $X(L)=0$ and, thus, be in global force balance in axial direction.

\section{Rotational stability}
\label{sec:approt}

We want to study the change in the torque due to an
infinitesimally small rotation $\mathsf{R}$ of the velocity and the
gravitational force vector. 
Geometrically, this is equivalent to rotating the capsule. A
change of the velocity boundary condition
$\vec{u}_0 \rightarrow \vec{u}_0' = \mathsf{R} \vec{u}_0$
leads to new hydrodynamic surface forces
$\vec{f} \rightarrow \vec{f}'$.
As we did not change the capsule we can employ the reciprocal theorem
\begin{equation}
 \int \mathrm{d}A\, \vec{u}\cdot \vec{f}' 
    = \int \mathrm{d}A\,\vec{u}_0'\cdot\vec{f}
\end{equation}
and find
$\vec{f}' =  (\mathsf{R}^T)^{-1} \vec{f} = \mathsf{R}\vec{f}$.
The new hydrodynamic torque is, thus,
\begin{equation}
\vec{T_u}' = \int\mathrm{d}A\, \left(\vec{r}\times \left(\mathsf{R}
\vec{f}\right)\right) 
 = \mathsf{R} \int\mathrm{d}A\, \left(\left(\mathsf{R^{-1}}\vec{r}\right)
\times \vec{f} \right) 
\text{.}
\end{equation}
We limit ourselves to a linear analysis, that is, we perform an infinitesimal
rotation around the $y$ axis by an angle $\mathrm{d}\alpha$, i.e.,
$\mathsf{R}^{\pm 1} = \mathsf{I}\pm \mathrm{d}\alpha \mathsf{J}_y$.
Here $\mathsf{I}$ is the unit matrix ($I_{ij}=\delta_{ij}$) and
$\mathsf{J}_y$ generates  a rotation around the $y$ axis with
$\left(J_y\right)_{ij} = \delta_{i3}\delta_{j1} - \delta_{i1}\delta_{j3}$. Then,
to linear order in $\mathrm{d}\alpha$,
\begin{equation}
\vec{e}_y\cdot \vec{T_u}'= -2\pi \mathrm{d}\alpha \int \mathrm{d}s\, r
\left(\frac{1}{2} r f_r +
(z-z_0) f_z \right) \text{.}
\end{equation}
 Here $z_0$ gives the pivot point of the rotation. Since we know that the
 centers of mass and hydrodynamic stress have to lie on the symmetry axis,
 studying rotations about points on this axis suffices. Likewise, we can
 calculate the gravitational contribution to the new torque, which is 
(again to linear order in $\mathrm{d}\alpha$) 
 \begin{equation}
  \vec{e}_y\cdot \vec{T_g}' = \pi \mathrm{d}\alpha g \int \mathrm{d}s\, r^2
 (z-z_0)\text{.}
\end{equation}
Equating these two gives the pivot point of the marginally stable infinitesimal
rotation, i.e., the center of hydrodynamic stress. 

This argument does not account for a change of the capsule
shape due to the altered stresses upon rotation, which means that a seemingly
unstable capsule might be stabilized axially by a ``wobbling'' deformation
leading to different hydrodynamic torques. 
In principle, this might also work the other way round (as for
semiflexible cylindrical rods \cite{Manghi2006}), but we think it 
is less relevant
here, because of the typically large radius of curvature at the lower apex.

\section{Pressure control}
\label{sec:pressure}

For volume control, we fixed $V_0 = 4\pi/3$ to the volume of the spherical 
rest shape and only explored the two parameters 
 $\tilde{E}_B$ and $g$. The pressure $p_0$ served as Lagrange parameter 
to achieve the fixed volume $V_0$. 
To get a better understanding of the variety of shapes that are in
principle possible, we study capsules with volumes other than that of
the reference shape  by controlling and changing 
the  Lagrange parameter $p_0$. 
Thus, the explorable parameter space now consists of three
 parameters $\tilde{E}_B$,  $g$, and $p_0$.
As for volume control, the sedimenting velocity $u$ is 
not a control parameter but determined from 
 demanding a force free capsule.

 With the  additional pressure parameter $p_0$, we  find 
 more sedimenting shapes to be stable (with different volumes), 
in particular, more classes of  buckled  shapes. 
We tried to find solutions at $\tilde{E}_B=0.01$ for pressures
$p_0 \in [-0.95,0.95]$ and $g \in [0,1]$. As a consequence of the quadratic
nature of the elastic law, see Eq.\ (\ref{eq:Hook}),  
there is no solution for
a resting capsule ($g=0$) with $p_0>1$ \cite{Knoche2011}.
For a cleaner comparison of different parameter sets (in particular 
for comparison of differently elongated shapes) it is helpful to not use
$p_0$ but the pressure at the center of mass, $p_{\text{eff}} = p_0 - \langle
z\rangle g$, as an actual control parameter because we use 
a coordinate system that fixes $z=0$
at the lower apex of the capsule. 

\begin{figure}[htb]
\includegraphics[width=0.95\linewidth]{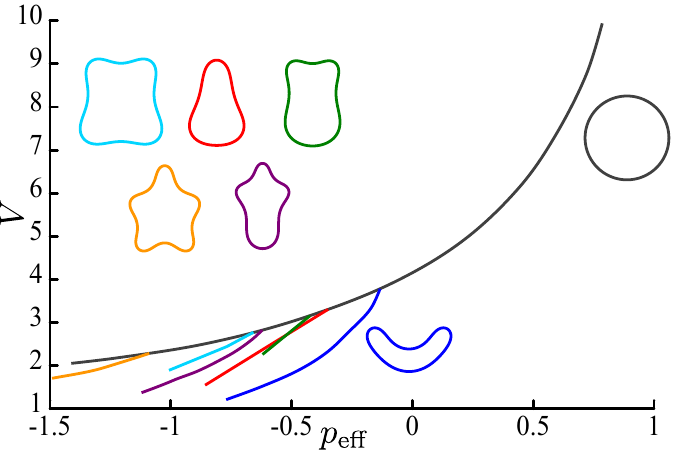}
\caption{  (Color online)
  Volume $V$ of the sedimenting elastic capsule as a function of the
  controlled pressure $p_\text{eff}$ at the center of mass at fixed bending
  rigidity $\tilde{E}_B=0.01$ and for shapes moving at a velocity $u\approx
  0.07$.  Each type of stable shape gives one pressure-volume branch. We
  identified the branches by the shape of the generatrices. We show one
  representative generatrix for each branch in the color of the branch. Please
  note that the generatrices are not drawn to scale.  The pseudospherical
  branch is in good quantitative agreement with the static pressure-volume
  relation of Eq.\ (\ref{eq:spherical}).  }
 \label{fig:pc}
\end{figure}

The possibility to explore a range of volumes allows for a greater
variety of shapes. 
In Fig.\ \ref{fig:pc}, we show the pressure-volume relation 
for stationary axisymmetric shapes
with a velocity close to $u=0.07$ (which determines for each shape 
a certain gravity $g$). Solutions at lower velocities are typically
easier to find, but very low velocities are rather atypical because one
ultimately sees the variety of static solutions \cite{Knoche2011}.
Each type of stable shape gives one pressure-volume branch 
in Fig.\ \ref{fig:pc}. Apart from pseudospherical, pear-shaped and 
the pair of weakly and strongly buckled solutions, we find 
 additional branches, which correspond to higher-order 
buckled shapes  with three, four, or five indentations.

As for fixed volume, the main contribution to the shape is due to static
pressure (differences). For the pseudospherical branch this can be seen by
comparing the volume (in reduced units) with the known pressure-volume
relation for a spherical capsule in the static ($g=0$) case
\begin{align}
 V_{\text{sphere}}(p)&= \frac{4\pi}{3} \left( \frac{2}{p} \left( 1-
\sqrt{1-p}\right)\right)^3 \,\text{,} 
\label{eq:spherical}
\end{align}
which fits our data rather well.

The higher-order buckled  shapes occur at small volumes or small 
effective pressure $p_{\text{eff}}$. 
We expect to find corresponding higher-order buckled shapes  also 
under volume control if we prescribe a corresponding  small volume
$V_0 \sim 1-2$.
But also for higher prescribed volumes such as 
$V_0= 4\pi/3\simeq 4.19$, these 
shapes could become stable at very high $g$: Higher 
volumes require a  larger pressure $p_{\text{eff}}$
and, thus, also higher deformation forces to 
induce buckling. Such higher deformation forces 
can be achieved at high $g$.

\begin{thebibliography}{53}%
\makeatletter
\providecommand \@ifxundefined [1]{%
 \@ifx{#1\undefined}
}%
\providecommand \@ifnum [1]{%
 \ifnum #1\expandafter \@firstoftwo
 \else \expandafter \@secondoftwo
 \fi
}%
\providecommand \@ifx [1]{%
 \ifx #1\expandafter \@firstoftwo
 \else \expandafter \@secondoftwo
 \fi
}%
\providecommand \natexlab [1]{#1}%
\providecommand \enquote  [1]{``#1''}%
\providecommand \bibnamefont  [1]{#1}%
\providecommand \bibfnamefont [1]{#1}%
\providecommand \citenamefont [1]{#1}%
\providecommand \href@noop [0]{\@secondoftwo}%
\providecommand \href [0]{\begingroup \@sanitize@url \@href}%
\providecommand \@href[1]{\@@startlink{#1}\@@href}%
\providecommand \@@href[1]{\endgroup#1\@@endlink}%
\providecommand \@sanitize@url [0]{\catcode `\\12\catcode `\$12\catcode
  `\&12\catcode `\#12\catcode `\^12\catcode `\_12\catcode `\%12\relax}%
\providecommand \@@startlink[1]{}%
\providecommand \@@endlink[0]{}%
\providecommand \url  [0]{\begingroup\@sanitize@url \@url }%
\providecommand \@url [1]{\endgroup\@href {#1}{\urlprefix }}%
\providecommand \urlprefix  [0]{URL }%
\providecommand \Eprint [0]{\href }%
\providecommand \doibase [0]{http://dx.doi.org/}%
\providecommand \selectlanguage [0]{\@gobble}%
\providecommand \bibinfo  [0]{\@secondoftwo}%
\providecommand \bibfield  [0]{\@secondoftwo}%
\providecommand \translation [1]{[#1]}%
\providecommand \BibitemOpen [0]{}%
\providecommand \bibitemStop [0]{}%
\providecommand \bibitemNoStop [0]{.\EOS\space}%
\providecommand \EOS [0]{\spacefactor3000\relax}%
\providecommand \BibitemShut  [1]{\csname bibitem#1\endcsname}%
\let\auto@bib@innerbib\@empty
\bibitem [{\citenamefont {Barth\`{e}s-Biesel}(2011)}]{Barthes-Biesel2011}%
  \BibitemOpen
  \bibfield  {author} {\bibinfo {author} {\bibfnamefont {D.}~\bibnamefont
  {Barth\`{e}s-Biesel}},\ }\href {\doibase 10.1016/j.cocis.2010.07.001}
  {\bibfield  {journal} {\bibinfo  {journal} {Curr. Opin. Colloid Interface
  Sci.}\ }\textbf {\bibinfo {volume} {16}},\ \bibinfo {pages} {3} (\bibinfo
  {year} {2011})}\BibitemShut {NoStop}%
\bibitem [{\citenamefont {Fedosov}\ \emph {et~al.}(2014)\citenamefont
  {Fedosov}, \citenamefont {Noguchi},\ and\ \citenamefont
  {Gompper}}]{Fedosov2014}%
  \BibitemOpen
  \bibfield  {author} {\bibinfo {author} {\bibfnamefont {D.~A.}\ \bibnamefont
  {Fedosov}}, \bibinfo {author} {\bibfnamefont {H.}~\bibnamefont {Noguchi}}, \
  and\ \bibinfo {author} {\bibfnamefont {G.}~\bibnamefont {Gompper}},\ }\href
  {\doibase 10.1007/s10237-013-0497-9} {\bibfield  {journal} {\bibinfo
  {journal} {Biomech. Model. Mechanobiol.}\ }\textbf {\bibinfo {volume} {13}},\
  \bibinfo {pages} {239} (\bibinfo {year} {2014})}\BibitemShut {NoStop}%
\bibitem [{\citenamefont {Freund}(2014)}]{Freund2014}%
  \BibitemOpen
  \bibfield  {author} {\bibinfo {author} {\bibfnamefont {J.~B.}\ \bibnamefont
  {Freund}},\ }\href {\doibase 10.1146/annurev-fluid-010313-141349} {\bibfield
  {journal} {\bibinfo  {journal} {Annu. Rev. Fluid Mech.}\ }\textbf {\bibinfo
  {volume} {46}},\ \bibinfo {pages} {67} (\bibinfo {year} {2014})}\BibitemShut
  {NoStop}%
\bibitem [{\citenamefont {Abreu}\ \emph {et~al.}(2014)\citenamefont {Abreu},
  \citenamefont {Levant}, \citenamefont {Steinberg},\ and\ \citenamefont
  {Seifert}}]{Abreu2014}%
  \BibitemOpen
  \bibfield  {author} {\bibinfo {author} {\bibfnamefont {D.}~\bibnamefont
  {Abreu}}, \bibinfo {author} {\bibfnamefont {M.}~\bibnamefont {Levant}},
  \bibinfo {author} {\bibfnamefont {V.}~\bibnamefont {Steinberg}}, \ and\
  \bibinfo {author} {\bibfnamefont {U.}~\bibnamefont {Seifert}},\ }\href
  {\doibase 10.1016/j.cis.2014.02.004} {\bibfield  {journal} {\bibinfo
  {journal} {Adv. Colloid Interface Sci.}\ }\textbf {\bibinfo {volume} {208}},\
  \bibinfo {pages} {129} (\bibinfo {year} {2014})}\BibitemShut {NoStop}%
\bibitem [{\citenamefont {Huang}\ \emph {et~al.}(2011)\citenamefont {Huang},
  \citenamefont {Abkarian},\ and\ \citenamefont {Viallat}}]{huang2011}%
  \BibitemOpen
  \bibfield  {author} {\bibinfo {author} {\bibfnamefont {Z.}~\bibnamefont
  {Huang}}, \bibinfo {author} {\bibfnamefont {M.}~\bibnamefont {Abkarian}}, \
  and\ \bibinfo {author} {\bibfnamefont {A.}~\bibnamefont {Viallat}},\
  }\href@noop {} {\bibfield  {journal} {\bibinfo  {journal} {New J. Phys.}\
  }\textbf {\bibinfo {volume} {13}},\ \bibinfo {pages} {035026} (\bibinfo
  {year} {2011})}\BibitemShut {NoStop}%
\bibitem [{\citenamefont {Clift}\ \emph {et~al.}(1978)\citenamefont {Clift},
  \citenamefont {Grace},\ and\ \citenamefont {Weber}}]{clift1978bubbles}%
  \BibitemOpen
  \bibfield  {author} {\bibinfo {author} {\bibfnamefont {R.}~\bibnamefont
  {Clift}}, \bibinfo {author} {\bibfnamefont {J.}~\bibnamefont {Grace}}, \ and\
  \bibinfo {author} {\bibfnamefont {M.}~\bibnamefont {Weber}},\ }\href
  {http://books.google.de/books?id=n8gRAQAAIAAJ} {\emph {\bibinfo {title}
  {Bubbles, Drops, and Particles}}}\ (\bibinfo  {publisher} {Academic Press,
  New York},\ \bibinfo {year} {1978})\BibitemShut {NoStop}%
\bibitem [{\citenamefont {Stone}(1994)}]{Stone1994}%
  \BibitemOpen
  \bibfield  {author} {\bibinfo {author} {\bibfnamefont {H.~A.}\ \bibnamefont
  {Stone}},\ }\href {\doibase 10.1146/annurev.fl.26.010194.000433} {\bibfield
  {journal} {\bibinfo  {journal} {Annu. Rev. Fluid Mech.}\ }\textbf {\bibinfo
  {volume} {26}},\ \bibinfo {pages} {65} (\bibinfo {year} {1994})}\BibitemShut
  {NoStop}%
\bibitem [{\citenamefont {Pozrikidis}(1992)}]{Pozrikidis1992}%
  \BibitemOpen
  \bibfield  {author} {\bibinfo {author} {\bibfnamefont {C.}~\bibnamefont
  {Pozrikidis}},\ }\href@noop {} {\emph {\bibinfo {title} {{Boundary Integral
  and Singularity Methods for Linearized Viscous Flow}}}}\ (\bibinfo
  {publisher} {Cambridge University Press, Cambridge},\ \bibinfo {year}
  {1992})\BibitemShut {NoStop}%
\bibitem [{\citenamefont {Pozrikidis}(2001)}]{Pozrikidis2001}%
  \BibitemOpen
  \bibfield  {author} {\bibinfo {author} {\bibfnamefont {C.}~\bibnamefont
  {Pozrikidis}},\ }\href {\doibase 10.1006/jcph.2000.6582} {\bibfield
  {journal} {\bibinfo  {journal} {J. Comput. Phys.}\ }\textbf {\bibinfo
  {volume} {169}},\ \bibinfo {pages} {250} (\bibinfo {year}
  {2001})}\BibitemShut {NoStop}%
\bibitem [{\citenamefont {Meier}(2000)}]{Meier2000}%
  \BibitemOpen
  \bibfield  {author} {\bibinfo {author} {\bibfnamefont {W.}~\bibnamefont
  {Meier}},\ }\href {\doibase 10.1039/a809106d} {\bibfield  {journal} {\bibinfo
   {journal} {Chem. Soc. Rev.}\ }\textbf {\bibinfo {volume} {29}},\ \bibinfo
  {pages} {295} (\bibinfo {year} {2000})}\BibitemShut {NoStop}%
\bibitem [{\citenamefont {Neubauer}\ \emph {et~al.}(2014)\citenamefont
  {Neubauer}, \citenamefont {Poehlmann},\ and\ \citenamefont
  {Fery}}]{Neubauer2013}%
  \BibitemOpen
  \bibfield  {author} {\bibinfo {author} {\bibfnamefont {M.~P.}\ \bibnamefont
  {Neubauer}}, \bibinfo {author} {\bibfnamefont {M.}~\bibnamefont {Poehlmann}},
  \ and\ \bibinfo {author} {\bibfnamefont {A.}~\bibnamefont {Fery}},\ }\href
  {\doibase 10.1016/j.cis.2013.11.016} {\bibfield  {journal} {\bibinfo
  {journal} {Adv. Colloid Interface Sci.}\ }\textbf {\bibinfo {volume} {207}},\
  \bibinfo {pages} {65} (\bibinfo {year} {2014})}\BibitemShut {NoStop}%
\bibitem [{\citenamefont {Pontani}\ \emph {et~al.}(2009)\citenamefont
  {Pontani}, \citenamefont {{Van Der Gucht}}, \citenamefont {Salbreux},
  \citenamefont {Heuvingh}, \citenamefont {Joanny},\ and\ \citenamefont
  {Sykes}}]{Pontani2009}%
  \BibitemOpen
  \bibfield  {author} {\bibinfo {author} {\bibfnamefont {L.~L.}\ \bibnamefont
  {Pontani}}, \bibinfo {author} {\bibfnamefont {J.}~\bibnamefont {{Van Der
  Gucht}}}, \bibinfo {author} {\bibfnamefont {G.}~\bibnamefont {Salbreux}},
  \bibinfo {author} {\bibfnamefont {J.}~\bibnamefont {Heuvingh}}, \bibinfo
  {author} {\bibfnamefont {J.~F.}\ \bibnamefont {Joanny}}, \ and\ \bibinfo
  {author} {\bibfnamefont {C.}~\bibnamefont {Sykes}},\ }\href {\doibase
  10.1016/j.bpj.2008.09.029} {\bibfield  {journal} {\bibinfo  {journal}
  {Biophys. J.}\ }\textbf {\bibinfo {volume} {96}},\ \bibinfo {pages} {192}
  (\bibinfo {year} {2009})}\BibitemShut {NoStop}%
\bibitem [{\citenamefont {Tsai}\ \emph {et~al.}(2011)\citenamefont {Tsai},
  \citenamefont {Stuhrmann},\ and\ \citenamefont {Koenderink}}]{Tsai2011}%
  \BibitemOpen
  \bibfield  {author} {\bibinfo {author} {\bibfnamefont {F.~C.}\ \bibnamefont
  {Tsai}}, \bibinfo {author} {\bibfnamefont {B.}~\bibnamefont {Stuhrmann}}, \
  and\ \bibinfo {author} {\bibfnamefont {G.~H.}\ \bibnamefont {Koenderink}},\
  }\href {\doibase 10.1021/la201604z} {\bibfield  {journal} {\bibinfo
  {journal} {Langmuir}\ }\textbf {\bibinfo {volume} {27}},\ \bibinfo {pages}
  {10061} (\bibinfo {year} {2011})}\BibitemShut {NoStop}%
\bibitem [{\citenamefont {Carvalho}\ \emph {et~al.}(2013)\citenamefont
  {Carvalho}, \citenamefont {Tsai}, \citenamefont {Lees}, \citenamefont
  {Voituiriez}, \citenamefont {Koenderink},\ and\ \citenamefont
  {Sykes}}]{Carvalho2013}%
  \BibitemOpen
  \bibfield  {author} {\bibinfo {author} {\bibfnamefont {K.}~\bibnamefont
  {Carvalho}}, \bibinfo {author} {\bibfnamefont {F.-C.~F.}\ \bibnamefont
  {Tsai}}, \bibinfo {author} {\bibfnamefont {E.}~\bibnamefont {Lees}}, \bibinfo
  {author} {\bibfnamefont {R.}~\bibnamefont {Voituiriez}}, \bibinfo {author}
  {\bibfnamefont {G.}~\bibnamefont {Koenderink}}, \ and\ \bibinfo {author}
  {\bibfnamefont {C.}~\bibnamefont {Sykes}},\ }\href {\doibase
  10.1073/pnas.1320015110} {\bibfield  {journal} {\bibinfo  {journal} {Proc.
  Natl. Acad. Sci. U.S.A.}\ }\textbf {\bibinfo {volume} {110}},\ \bibinfo
  {pages} {16456} (\bibinfo {year} {2013})}\BibitemShut {NoStop}%
\bibitem [{\citenamefont {Sch\"{a}fer}\ \emph {et~al.}(2013)\citenamefont
  {Sch\"{a}fer}, \citenamefont {Kliesch},\ and\ \citenamefont
  {Janshoff}}]{Schaefer2013}%
  \BibitemOpen
  \bibfield  {author} {\bibinfo {author} {\bibfnamefont {E.}~\bibnamefont
  {Sch\"{a}fer}}, \bibinfo {author} {\bibfnamefont {T.~T.}\ \bibnamefont
  {Kliesch}}, \ and\ \bibinfo {author} {\bibfnamefont {A.}~\bibnamefont
  {Janshoff}},\ }\href {\doibase 10.1021/la401969t} {\bibfield  {journal}
  {\bibinfo  {journal} {Langmuir}\ }\textbf {\bibinfo {volume} {29}},\ \bibinfo
  {pages} {10463} (\bibinfo {year} {2013})}\BibitemShut {NoStop}%
\bibitem [{\citenamefont {L\'{o}pez-Montero}\ \emph {et~al.}(2012)\citenamefont
  {L\'{o}pez-Montero}, \citenamefont {Rodr\'{\i}guez-Garc\'{\i}a},\ and\
  \citenamefont {Monroy}}]{Lopez-Montero2012}%
  \BibitemOpen
  \bibfield  {author} {\bibinfo {author} {\bibfnamefont {I.}~\bibnamefont
  {L\'{o}pez-Montero}}, \bibinfo {author} {\bibfnamefont {R.}~\bibnamefont
  {Rodr\'{\i}guez-Garc\'{\i}a}}, \ and\ \bibinfo {author} {\bibfnamefont
  {F.}~\bibnamefont {Monroy}},\ }\href {\doibase 10.1021/jz300377q} {\bibfield
  {journal} {\bibinfo  {journal} {J. Phys. Chem. Lett.}\ }\textbf {\bibinfo
  {volume} {3}},\ \bibinfo {pages} {1583} (\bibinfo {year} {2012})}\BibitemShut
  {NoStop}%
\bibitem [{\citenamefont {Saha}\ \emph {et~al.}(2013)\citenamefont {Saha},
  \citenamefont {Mondal}, \citenamefont {Biswas}, \citenamefont {Chakraborty},
  \citenamefont {Jana},\ and\ \citenamefont {Ghosh}}]{Saha2013}%
  \BibitemOpen
  \bibfield  {author} {\bibinfo {author} {\bibfnamefont {A.}~\bibnamefont
  {Saha}}, \bibinfo {author} {\bibfnamefont {G.}~\bibnamefont {Mondal}},
  \bibinfo {author} {\bibfnamefont {A.}~\bibnamefont {Biswas}}, \bibinfo
  {author} {\bibfnamefont {I.}~\bibnamefont {Chakraborty}}, \bibinfo {author}
  {\bibfnamefont {B.}~\bibnamefont {Jana}}, \ and\ \bibinfo {author}
  {\bibfnamefont {S.}~\bibnamefont {Ghosh}},\ }\href {\doibase
  10.1039/c3cc41287c} {\bibfield  {journal} {\bibinfo  {journal} {Chem.
  Commun.}\ }\textbf {\bibinfo {volume} {49}},\ \bibinfo {pages} {6119}
  (\bibinfo {year} {2013})}\BibitemShut {NoStop}%
\bibitem [{\citenamefont {Maeda}\ \emph {et~al.}(2012)\citenamefont {Maeda},
  \citenamefont {Nakadai}, \citenamefont {Shin}, \citenamefont {Uryu},
  \citenamefont {Noireaux},\ and\ \citenamefont {Libchaber}}]{Maeda2012}%
  \BibitemOpen
  \bibfield  {author} {\bibinfo {author} {\bibfnamefont {Y.~T.}\ \bibnamefont
  {Maeda}}, \bibinfo {author} {\bibfnamefont {T.}~\bibnamefont {Nakadai}},
  \bibinfo {author} {\bibfnamefont {J.}~\bibnamefont {Shin}}, \bibinfo {author}
  {\bibfnamefont {K.}~\bibnamefont {Uryu}}, \bibinfo {author} {\bibfnamefont
  {V.}~\bibnamefont {Noireaux}}, \ and\ \bibinfo {author} {\bibfnamefont
  {A.}~\bibnamefont {Libchaber}},\ }\href {\doibase 10.1021/sb200003v}
  {\bibfield  {journal} {\bibinfo  {journal} {ACS Synth Biol.}\ }\textbf
  {\bibinfo {volume} {1}},\ \bibinfo {pages} {53} (\bibinfo {year}
  {2012})}\BibitemShut {NoStop}%
\bibitem [{\citenamefont {Vinogradova}\ \emph {et~al.}(2006)\citenamefont
  {Vinogradova}, \citenamefont {Lebedeva},\ and\ \citenamefont
  {Kim}}]{Vinogradova2006}%
  \BibitemOpen
  \bibfield  {author} {\bibinfo {author} {\bibfnamefont {O.~I.}\ \bibnamefont
  {Vinogradova}}, \bibinfo {author} {\bibfnamefont {O.~V.}\ \bibnamefont
  {Lebedeva}}, \ and\ \bibinfo {author} {\bibfnamefont {B.-S.}\ \bibnamefont
  {Kim}},\ }\href {\doibase 10.1146/annurev.matsci.36.011205.123733} {\bibfield
   {journal} {\bibinfo  {journal} {Annu. Rev. Mater. Res.}\ }\textbf {\bibinfo
  {volume} {36}},\ \bibinfo {pages} {143} (\bibinfo {year} {2006})}\BibitemShut
  {NoStop}%
\bibitem [{\citenamefont {Lidmar}\ \emph {et~al.}(2003)\citenamefont {Lidmar},
  \citenamefont {Mirny},\ and\ \citenamefont {Nelson}}]{Lidmar2003}%
  \BibitemOpen
  \bibfield  {author} {\bibinfo {author} {\bibfnamefont {J.}~\bibnamefont
  {Lidmar}}, \bibinfo {author} {\bibfnamefont {L.}~\bibnamefont {Mirny}}, \
  and\ \bibinfo {author} {\bibfnamefont {D.~R.}\ \bibnamefont {Nelson}},\
  }\href {\doibase 10.1103/PhysRevE.68.051910} {\bibfield  {journal} {\bibinfo
  {journal} {Phys. Rev. E}\ }\textbf {\bibinfo {volume} {68}},\ \bibinfo
  {pages} {051910} (\bibinfo {year} {2003})}\BibitemShut {NoStop}%
\bibitem [{\citenamefont {Mukhopadhyay}\ \emph {et~al.}(2002)\citenamefont
  {Mukhopadhyay}, \citenamefont {{Lim H W}},\ and\ \citenamefont
  {Wortis}}]{Mukhopadhyay2002}%
  \BibitemOpen
  \bibfield  {author} {\bibinfo {author} {\bibfnamefont {R.}~\bibnamefont
  {Mukhopadhyay}}, \bibinfo {author} {\bibfnamefont {G.}~\bibnamefont {{Lim H
  W}}}, \ and\ \bibinfo {author} {\bibfnamefont {M.}~\bibnamefont {Wortis}},\
  }\href {\doibase 10.1016/S0006-3495(02)75527-6} {\bibfield  {journal}
  {\bibinfo  {journal} {Biophys. J.}\ }\textbf {\bibinfo {volume} {82}},\
  \bibinfo {pages} {1756} (\bibinfo {year} {2002})},\ \Eprint
  {http://arxiv.org/abs/0108122} {arXiv:0108122 [cond-mat]} \BibitemShut
  {NoStop}%
\bibitem [{\citenamefont {{Lim H W}}\ \emph {et~al.}(2002)\citenamefont {{Lim H
  W}}, \citenamefont {Wortis},\ and\ \citenamefont {Mukhopadhyay}}]{LimHW2002}%
  \BibitemOpen
  \bibfield  {author} {\bibinfo {author} {\bibfnamefont {G.}~\bibnamefont {{Lim
  H W}}}, \bibinfo {author} {\bibfnamefont {M.}~\bibnamefont {Wortis}}, \ and\
  \bibinfo {author} {\bibfnamefont {R.}~\bibnamefont {Mukhopadhyay}},\ }\href
  {\doibase 10.1073/pnas.202617299} {\bibfield  {journal} {\bibinfo  {journal}
  {Proc. Natl. Acad. Sci. U.S.A.}\ }\textbf {\bibinfo {volume} {99}},\ \bibinfo
  {pages} {16766} (\bibinfo {year} {2002})}\BibitemShut {NoStop}%
\bibitem [{\citenamefont {Boedec}\ \emph {et~al.}(2011)\citenamefont {Boedec},
  \citenamefont {Leonetti},\ and\ \citenamefont {Jaeger}}]{Boedec2011a}%
  \BibitemOpen
  \bibfield  {author} {\bibinfo {author} {\bibfnamefont {G.}~\bibnamefont
  {Boedec}}, \bibinfo {author} {\bibfnamefont {M.}~\bibnamefont {Leonetti}}, \
  and\ \bibinfo {author} {\bibfnamefont {M.}~\bibnamefont {Jaeger}},\
  }\href@noop {} {\bibfield  {journal} {\bibinfo  {journal} {J. Comput. Phys.}\
  }\textbf {\bibinfo {volume} {230}},\ \bibinfo {pages} {1020} (\bibinfo {year}
  {2011})}\BibitemShut {NoStop}%
\bibitem [{\citenamefont {Boedec}\ \emph {et~al.}(2012)\citenamefont {Boedec},
  \citenamefont {Jaeger},\ and\ \citenamefont {Leonetti}}]{Boedec2012}%
  \BibitemOpen
  \bibfield  {author} {\bibinfo {author} {\bibfnamefont {G.}~\bibnamefont
  {Boedec}}, \bibinfo {author} {\bibfnamefont {M.}~\bibnamefont {Jaeger}}, \
  and\ \bibinfo {author} {\bibfnamefont {M.}~\bibnamefont {Leonetti}},\ }\href
  {\doibase 10.1017/jfm.2011.427} {\bibfield  {journal} {\bibinfo  {journal}
  {J. Fluid Mech.}\ }\textbf {\bibinfo {volume} {690}},\ \bibinfo {pages} {227}
  (\bibinfo {year} {2012})}\BibitemShut {NoStop}%
\bibitem [{\citenamefont {Boedec}\ \emph {et~al.}(2013)\citenamefont {Boedec},
  \citenamefont {Jaeger},\ and\ \citenamefont {Leonetti}}]{Boedec2013}%
  \BibitemOpen
  \bibfield  {author} {\bibinfo {author} {\bibfnamefont {G.}~\bibnamefont
  {Boedec}}, \bibinfo {author} {\bibfnamefont {M.}~\bibnamefont {Jaeger}}, \
  and\ \bibinfo {author} {\bibfnamefont {M.}~\bibnamefont {Leonetti}},\ }\href
  {\doibase 10.1103/PhysRevE.88.010702} {\bibfield  {journal} {\bibinfo
  {journal} {Phys. Rev. E}\ }\textbf {\bibinfo {volume} {88}},\ \bibinfo
  {pages} {010702} (\bibinfo {year} {2013})}\BibitemShut {NoStop}%
\bibitem [{\citenamefont {{Rey Su\'{a}rez}}\ \emph {et~al.}(2013)\citenamefont
  {{Rey Su\'{a}rez}}, \citenamefont {Leidy}, \citenamefont {T\'{e}llez},
  \citenamefont {Gay},\ and\ \citenamefont {Gonzalez-Mancera}}]{ReySuarez2013}%
  \BibitemOpen
  \bibfield  {author} {\bibinfo {author} {\bibfnamefont {I.}~\bibnamefont {{Rey
  Su\'{a}rez}}}, \bibinfo {author} {\bibfnamefont {C.}~\bibnamefont {Leidy}},
  \bibinfo {author} {\bibfnamefont {G.}~\bibnamefont {T\'{e}llez}}, \bibinfo
  {author} {\bibfnamefont {G.}~\bibnamefont {Gay}}, \ and\ \bibinfo {author}
  {\bibfnamefont {A.}~\bibnamefont {Gonzalez-Mancera}},\ }\href {\doibase
  10.1371/journal.pone.0068309} {\bibfield  {journal} {\bibinfo  {journal}
  {PLoS ONE}\ }\textbf {\bibinfo {volume} {8}},\ \bibinfo {pages} {e68309}
  (\bibinfo {year} {2013})}\BibitemShut {NoStop}%
\bibitem [{\citenamefont {Corry}\ and\ \citenamefont
  {Meiselman}(1978{\natexlab{a}})}]{Corry1978}%
  \BibitemOpen
  \bibfield  {author} {\bibinfo {author} {\bibfnamefont {W.~D.}\ \bibnamefont
  {Corry}}\ and\ \bibinfo {author} {\bibfnamefont {H.~J.}\ \bibnamefont
  {Meiselman}},\ }\href {\doibase 10.1016/S0006-3495(78)85506-4} {\bibfield
  {journal} {\bibinfo  {journal} {Biophys. J.}\ }\textbf {\bibinfo {volume}
  {21}},\ \bibinfo {pages} {19} (\bibinfo {year}
  {1978}{\natexlab{a}})}\BibitemShut {NoStop}%
\bibitem [{\citenamefont {Corry}\ and\ \citenamefont
  {Meiselman}(1978{\natexlab{b}})}]{Corry1978a}%
  \BibitemOpen
  \bibfield  {author} {\bibinfo {author} {\bibfnamefont {W.~D.}\ \bibnamefont
  {Corry}}\ and\ \bibinfo {author} {\bibfnamefont {H.~J.}\ \bibnamefont
  {Meiselman}},\ }\href {http://www.ncbi.nlm.nih.gov/pubmed/415773} {\bibfield
  {journal} {\bibinfo  {journal} {Blood}\ }\textbf {\bibinfo {volume} {51}},\
  \bibinfo {pages} {693} (\bibinfo {year} {1978}{\natexlab{b}})}\BibitemShut
  {NoStop}%
\bibitem [{\citenamefont {Hoffman}\ and\ \citenamefont
  {Inou\'{e}}(2006)}]{Hoffman2006}%
  \BibitemOpen
  \bibfield  {author} {\bibinfo {author} {\bibfnamefont {J.~F.}\ \bibnamefont
  {Hoffman}}\ and\ \bibinfo {author} {\bibfnamefont {S.}~\bibnamefont
  {Inou\'{e}}},\ }\href {\doibase 10.1073/pnas.0510884103} {\bibfield
  {journal} {\bibinfo  {journal} {Proc. Natl. Acad. Sci. U.S.A.}\ }\textbf
  {\bibinfo {volume} {103}},\ \bibinfo {pages} {2971} (\bibinfo {year}
  {2006})}\BibitemShut {NoStop}%
\bibitem [{\citenamefont {Peltom{\"a}ki}\ and\ \citenamefont
  {Gompper}(2013)}]{Peltomaki2013}%
  \BibitemOpen
  \bibfield  {author} {\bibinfo {author} {\bibfnamefont {M.}~\bibnamefont
  {Peltom{\"a}ki}}\ and\ \bibinfo {author} {\bibfnamefont {G.}~\bibnamefont
  {Gompper}},\ }\href@noop {} {\bibfield  {journal} {\bibinfo  {journal} {Soft
  Matter}\ }\textbf {\bibinfo {volume} {9}},\ \bibinfo {pages} {8346} (\bibinfo
  {year} {2013})}\BibitemShut {NoStop}%
\bibitem [{\citenamefont {Noguchi}\ and\ \citenamefont
  {Gompper}(2005)}]{Noguchi2005}%
  \BibitemOpen
  \bibfield  {author} {\bibinfo {author} {\bibfnamefont {H.}~\bibnamefont
  {Noguchi}}\ and\ \bibinfo {author} {\bibfnamefont {G.}~\bibnamefont
  {Gompper}},\ }\href {\doibase 10.1073/pnas.0504243102} {\bibfield  {journal}
  {\bibinfo  {journal} {Proc. Natl. Acad. Sci. U.S.A.}\ }\textbf {\bibinfo
  {volume} {102}},\ \bibinfo {pages} {14159} (\bibinfo {year}
  {2005})}\BibitemShut {NoStop}%
\bibitem [{\citenamefont {Fedosov}\ \emph {et~al.}(2011)\citenamefont
  {Fedosov}, \citenamefont {Pan}, \citenamefont {Caswell}, \citenamefont
  {Gompper},\ and\ \citenamefont {Karniadakis}}]{Fedosov2011}%
  \BibitemOpen
  \bibfield  {author} {\bibinfo {author} {\bibfnamefont {D.~A.}\ \bibnamefont
  {Fedosov}}, \bibinfo {author} {\bibfnamefont {W.}~\bibnamefont {Pan}},
  \bibinfo {author} {\bibfnamefont {B.}~\bibnamefont {Caswell}}, \bibinfo
  {author} {\bibfnamefont {G.}~\bibnamefont {Gompper}}, \ and\ \bibinfo
  {author} {\bibfnamefont {G.~E.}\ \bibnamefont {Karniadakis}},\ }\href
  {\doibase 10.1073/pnas.1101210108} {\bibfield  {journal} {\bibinfo  {journal}
  {Proc. Natl. Acad. Sci. U.S.A.}\ }\textbf {\bibinfo {volume} {108}},\
  \bibinfo {pages} {11772} (\bibinfo {year} {2011})}\BibitemShut {NoStop}%
\bibitem [{\citenamefont {Dupin}\ \emph {et~al.}(2007)\citenamefont {Dupin},
  \citenamefont {Halliday}, \citenamefont {Care}, \citenamefont {Alboul},\ and\
  \citenamefont {Munn}}]{Dupin2007}%
  \BibitemOpen
  \bibfield  {author} {\bibinfo {author} {\bibfnamefont {M.~M.}\ \bibnamefont
  {Dupin}}, \bibinfo {author} {\bibfnamefont {I.}~\bibnamefont {Halliday}},
  \bibinfo {author} {\bibfnamefont {C.~M.}\ \bibnamefont {Care}}, \bibinfo
  {author} {\bibfnamefont {L.}~\bibnamefont {Alboul}}, \ and\ \bibinfo {author}
  {\bibfnamefont {L.~L.}\ \bibnamefont {Munn}},\ }\href {\doibase
  10.1103/PhysRevE.75.066707} {\bibfield  {journal} {\bibinfo  {journal} {Phys.
  Rev. E}\ }\textbf {\bibinfo {volume} {75}},\ \bibinfo {pages} {066707}
  (\bibinfo {year} {2007})}\BibitemShut {NoStop}%
\bibitem [{\citenamefont {Happel}\ and\ \citenamefont
  {Brenner}(1983)}]{HappelBrenner1983}%
  \BibitemOpen
  \bibfield  {author} {\bibinfo {author} {\bibfnamefont {J.}~\bibnamefont
  {Happel}}\ and\ \bibinfo {author} {\bibfnamefont {H.}~\bibnamefont
  {Brenner}},\ }\href@noop {} {\emph {\bibinfo {title} {{Low Reynolds Number
  Hydrodynamics: With Special Applications to Particulate Media}}}},\
  Vol.~\bibinfo {volume} {1}\ (\bibinfo  {publisher} {Springer, Berlin},\
  \bibinfo {year} {1983})\BibitemShut {NoStop}%
\bibitem [{\citenamefont {Lauga}\ and\ \citenamefont
  {Powers}(2009)}]{Lauga2009}%
  \BibitemOpen
  \bibfield  {author} {\bibinfo {author} {\bibfnamefont {E.}~\bibnamefont
  {Lauga}}\ and\ \bibinfo {author} {\bibfnamefont {T.~R.}\ \bibnamefont
  {Powers}},\ }\href {\doibase 10.1088/0034-4885/72/9/096601} {\bibfield
  {journal} {\bibinfo  {journal} {Rep. Prog. Phys.}\ }\textbf {\bibinfo
  {volume} {72}},\ \bibinfo {pages} {096601} (\bibinfo {year}
  {2009})}\BibitemShut {NoStop}%
\bibitem [{\citenamefont {Degen}(2014)}]{degen2014}%
  \BibitemOpen
  \bibfield  {author} {\bibinfo {author} {\bibfnamefont {P.}~\bibnamefont
  {Degen}},\ }\href@noop {} {\bibfield  {journal} {\bibinfo  {journal} {Curr.
  Opin. Colloid Interface Sci.}\ }\textbf {\bibinfo {volume} {19}},\ \bibinfo
  {pages} {611} (\bibinfo {year} {2014})}\BibitemShut {NoStop}%
\bibitem [{\citenamefont {Knoche}\ and\ \citenamefont
  {Kierfeld}(2011)}]{Knoche2011}%
  \BibitemOpen
  \bibfield  {author} {\bibinfo {author} {\bibfnamefont {S.}~\bibnamefont
  {Knoche}}\ and\ \bibinfo {author} {\bibfnamefont {J.}~\bibnamefont
  {Kierfeld}},\ }\href {\doibase 10.1103/PhysRevE.84.046608} {\bibfield
  {journal} {\bibinfo  {journal} {Phys. Rev. E}\ }\textbf {\bibinfo {volume}
  {84}},\ \bibinfo {pages} {046608} (\bibinfo {year} {2011})}\BibitemShut
  {NoStop}%
\bibitem [{\citenamefont {Knoche}\ \emph {et~al.}(2013)\citenamefont {Knoche},
  \citenamefont {Vella}, \citenamefont {Aumaitre}, \citenamefont {Degen},
  \citenamefont {Rehage}, \citenamefont {Cicuta},\ and\ \citenamefont
  {Kierfeld}}]{Knoche2013}%
  \BibitemOpen
  \bibfield  {author} {\bibinfo {author} {\bibfnamefont {S.}~\bibnamefont
  {Knoche}}, \bibinfo {author} {\bibfnamefont {D.}~\bibnamefont {Vella}},
  \bibinfo {author} {\bibfnamefont {E.}~\bibnamefont {Aumaitre}}, \bibinfo
  {author} {\bibfnamefont {P.}~\bibnamefont {Degen}}, \bibinfo {author}
  {\bibfnamefont {H.}~\bibnamefont {Rehage}}, \bibinfo {author} {\bibfnamefont
  {P.}~\bibnamefont {Cicuta}}, \ and\ \bibinfo {author} {\bibfnamefont
  {J.}~\bibnamefont {Kierfeld}},\ }\href {\doibase 10.1021/la402322g}
  {\bibfield  {journal} {\bibinfo  {journal} {Langmuir}\ }\textbf {\bibinfo
  {volume} {29}},\ \bibinfo {pages} {12463} (\bibinfo {year}
  {2013})}\BibitemShut {NoStop}%
\bibitem [{\citenamefont {Libai}\ and\ \citenamefont
  {Simmonds}(1998)}]{LibaiSimmonds1998}%
  \BibitemOpen
  \bibfield  {author} {\bibinfo {author} {\bibfnamefont {A.}~\bibnamefont
  {Libai}}\ and\ \bibinfo {author} {\bibfnamefont {J.}~\bibnamefont
  {Simmonds}},\ }\href@noop {} {\emph {\bibinfo {title} {{The Nonlinear Theory
  of Elastic Shells}}}}\ (\bibinfo  {publisher} {Cambridge University Press,
  Cambridge},\ \bibinfo {year} {1998})\BibitemShut {NoStop}%
\bibitem [{\citenamefont {Pozrikidis}(2003)}]{Pozrikidis2003}%
  \BibitemOpen
  \bibfield  {author} {\bibinfo {author} {\bibfnamefont {C.}~\bibnamefont
  {Pozrikidis}},\ }\href@noop {} {\emph {\bibinfo {title} {{Modeling and
  Simulation of Capsules and Biological Cells}}}}\ (\bibinfo  {publisher} {CRC
  Press, Boca Raton, FL},\ \bibinfo {year} {2003})\BibitemShut {NoStop}%
\bibitem [{\citenamefont {Landau}\ and\ \citenamefont {Lifshitz}(1986)}]{LL7}%
  \BibitemOpen
  \bibfield  {author} {\bibinfo {author} {\bibfnamefont {L.}~\bibnamefont
  {Landau}}\ and\ \bibinfo {author} {\bibfnamefont {E.}~\bibnamefont
  {Lifshitz}},\ }\href@noop {} {\emph {\bibinfo {title} {Theory of
  Elasticity}}},\ Vol.~\bibinfo {volume} {7}\ (\bibinfo  {publisher} {Pergamon,
  New York},\ \bibinfo {year} {1986})\BibitemShut {NoStop}%
\bibitem [{\citenamefont {Pogorelov}(1988)}]{pogorelov1988}%
  \BibitemOpen
  \bibfield  {author} {\bibinfo {author} {\bibfnamefont {A.~V.}\ \bibnamefont
  {Pogorelov}},\ }\href@noop {} {\emph {\bibinfo {title} {Bendings of surfaces
  and stability of shells}}},\ Vol.~\bibinfo {volume} {72}\ (\bibinfo
  {publisher} {American Mathematical Soc., Providence},\ \bibinfo {year}
  {1988})\BibitemShut {NoStop}%
\bibitem [{\citenamefont {Knoche}\ and\ \citenamefont
  {Kierfeld}(2014)}]{Knoche2014}%
  \BibitemOpen
  \bibfield  {author} {\bibinfo {author} {\bibfnamefont {S.}~\bibnamefont
  {Knoche}}\ and\ \bibinfo {author} {\bibfnamefont {J.}~\bibnamefont
  {Kierfeld}},\ }\href {\doibase 10.1039/c4sm01205d} {\bibfield  {journal}
  {\bibinfo  {journal} {Soft Matter}\ }\textbf {\bibinfo {volume} {10}},\
  \bibinfo {pages} {8358} (\bibinfo {year} {2014})}\BibitemShut {NoStop}%
\bibitem [{\citenamefont {Planken}\ and\ \citenamefont
  {C\"{o}lfen}(2010)}]{Planken2010}%
  \BibitemOpen
  \bibfield  {author} {\bibinfo {author} {\bibfnamefont {K.~L.}\ \bibnamefont
  {Planken}}\ and\ \bibinfo {author} {\bibfnamefont {H.}~\bibnamefont
  {C\"{o}lfen}},\ }\href {\doibase 10.1039/c0nr00215a} {\bibfield  {journal}
  {\bibinfo  {journal} {Nanoscale}\ }\textbf {\bibinfo {volume} {2}},\ \bibinfo
  {pages} {1849} (\bibinfo {year} {2010})}\BibitemShut {NoStop}%
\bibitem [{\citenamefont {ten Hagen}\ \emph {et~al.}(2014)\citenamefont {ten
  Hagen}, \citenamefont {K\"{u}mmel}, \citenamefont {Wittkowski}, \citenamefont
  {Takagi}, \citenamefont {L\"{o}wen},\ and\ \citenamefont
  {Bechinger}}]{TenHagen2014}%
  \BibitemOpen
  \bibfield  {author} {\bibinfo {author} {\bibfnamefont {B.}~\bibnamefont {ten
  Hagen}}, \bibinfo {author} {\bibfnamefont {F.}~\bibnamefont {K\"{u}mmel}},
  \bibinfo {author} {\bibfnamefont {R.}~\bibnamefont {Wittkowski}}, \bibinfo
  {author} {\bibfnamefont {D.}~\bibnamefont {Takagi}}, \bibinfo {author}
  {\bibfnamefont {H.}~\bibnamefont {L\"{o}wen}}, \ and\ \bibinfo {author}
  {\bibfnamefont {C.}~\bibnamefont {Bechinger}},\ }\href {\doibase
  10.1038/ncomms5829} {\bibfield  {journal} {\bibinfo  {journal} {Nat.
  Commun.}\ }\textbf {\bibinfo {volume} {5}},\ \bibinfo {pages} {4829}
  (\bibinfo {year} {2014})}\BibitemShut {NoStop}%
\bibitem [{\citenamefont {Lenormand}\ \emph {et~al.}(2001)\citenamefont
  {Lenormand}, \citenamefont {H\'{e}non}, \citenamefont {Richert},
  \citenamefont {Sim\'{e}on},\ and\ \citenamefont {Gallet}}]{Lenormand2001}%
  \BibitemOpen
  \bibfield  {author} {\bibinfo {author} {\bibfnamefont {G.}~\bibnamefont
  {Lenormand}}, \bibinfo {author} {\bibfnamefont {S.}~\bibnamefont
  {H\'{e}non}}, \bibinfo {author} {\bibfnamefont {a.}~\bibnamefont {Richert}},
  \bibinfo {author} {\bibfnamefont {J.}~\bibnamefont {Sim\'{e}on}}, \ and\
  \bibinfo {author} {\bibfnamefont {F.}~\bibnamefont {Gallet}},\ }\href
  {\doibase 10.1016/S0006-3495(01)75678-0} {\bibfield  {journal} {\bibinfo
  {journal} {Biophys. J.}\ }\textbf {\bibinfo {volume} {81}},\ \bibinfo {pages}
  {43} (\bibinfo {year} {2001})}\BibitemShut {NoStop}%
\bibitem [{\citenamefont {Ebbens}\ and\ \citenamefont
  {Howse}(2010)}]{ebbens2010}%
  \BibitemOpen
  \bibfield  {author} {\bibinfo {author} {\bibfnamefont {S.~J.}\ \bibnamefont
  {Ebbens}}\ and\ \bibinfo {author} {\bibfnamefont {J.~R.}\ \bibnamefont
  {Howse}},\ }\href@noop {} {\bibfield  {journal} {\bibinfo  {journal} {Soft
  Matter}\ }\textbf {\bibinfo {volume} {6}},\ \bibinfo {pages} {726} (\bibinfo
  {year} {2010})}\BibitemShut {NoStop}%
\bibitem [{\citenamefont {Dasgupta}\ and\ \citenamefont
  {Dimova}(2014)}]{dasgupta2014}%
  \BibitemOpen
  \bibfield  {author} {\bibinfo {author} {\bibfnamefont {R.}~\bibnamefont
  {Dasgupta}}\ and\ \bibinfo {author} {\bibfnamefont {R.}~\bibnamefont
  {Dimova}},\ }\href@noop {} {\bibfield  {journal} {\bibinfo  {journal} {J.
  Phys. D: Appl. Phys.}\ }\textbf {\bibinfo {volume} {47}},\ \bibinfo {pages}
  {282001} (\bibinfo {year} {2014})}\BibitemShut {NoStop}%
\bibitem [{\citenamefont {Lipowsky}\ \emph {et~al.}(2005)\citenamefont
  {Lipowsky}, \citenamefont {Brinkmann}, \citenamefont {Dimova}, \citenamefont
  {Franke}, \citenamefont {Kierfeld},\ and\ \citenamefont
  {Zhang}}]{Lipowsky2005}%
  \BibitemOpen
  \bibfield  {author} {\bibinfo {author} {\bibfnamefont {R.}~\bibnamefont
  {Lipowsky}}, \bibinfo {author} {\bibfnamefont {M.}~\bibnamefont {Brinkmann}},
  \bibinfo {author} {\bibfnamefont {R.}~\bibnamefont {Dimova}}, \bibinfo
  {author} {\bibfnamefont {T.}~\bibnamefont {Franke}}, \bibinfo {author}
  {\bibfnamefont {J.}~\bibnamefont {Kierfeld}}, \ and\ \bibinfo {author}
  {\bibfnamefont {X.}~\bibnamefont {Zhang}},\ }\href {\doibase
  10.1088/0953-8984/17/9/015} {\bibfield  {journal} {\bibinfo  {journal} {J.
  Phys.: Condens. Matter}\ }\textbf {\bibinfo {volume} {17}},\ \bibinfo {pages}
  {S537} (\bibinfo {year} {2005})}\BibitemShut {NoStop}%
\bibitem [{\citenamefont {Blake}(1971)}]{blake1971}%
  \BibitemOpen
  \bibfield  {author} {\bibinfo {author} {\bibfnamefont {J.}~\bibnamefont
  {Blake}},\ }\href@noop {} {\bibfield  {journal} {\bibinfo  {journal} {J. of
  Fluid Mech.}\ }\textbf {\bibinfo {volume} {46}},\ \bibinfo {pages} {199}
  (\bibinfo {year} {1971})}\BibitemShut {NoStop}%
\bibitem [{\citenamefont {Abramowitz}\ and\ \citenamefont
  {Stegun}(1972)}]{Abramowitz1972}%
  \BibitemOpen
  \bibfield  {author} {\bibinfo {author} {\bibfnamefont {M.}~\bibnamefont
  {Abramowitz}}\ and\ \bibinfo {author} {\bibfnamefont {I.~A.}\ \bibnamefont
  {Stegun}},\ }\href@noop {} {\emph {\bibinfo {title} {{Handbook of
  Mathematical Functions: With Formulas, Graphs, and Mathematical Tables}}}},\
  \bibinfo {number} {55}\ (\bibinfo  {publisher} {Dover Publications, New
  York},\ \bibinfo {year} {1972})\BibitemShut {NoStop}%
\bibitem [{\citenamefont {Perrin}(1936)}]{Perrin}%
  \BibitemOpen
  \bibfield  {author} {\bibinfo {author} {\bibfnamefont {F.}~\bibnamefont
  {Perrin}},\ }\href {\doibase 10.1051/jphysrad:01936007010100} {\bibfield
  {journal} {\bibinfo  {journal} {J. Phys. Radium}\ }\textbf {\bibinfo {volume}
  {7}},\ \bibinfo {pages} {1} (\bibinfo {year} {1936})}\BibitemShut {NoStop}%
\bibitem [{\citenamefont {Manghi}\ \emph {et~al.}(2006)\citenamefont {Manghi},
  \citenamefont {Schlagberger}, \citenamefont {Kim},\ and\ \citenamefont
  {Netz}}]{Manghi2006}%
  \BibitemOpen
  \bibfield  {author} {\bibinfo {author} {\bibfnamefont {M.}~\bibnamefont
  {Manghi}}, \bibinfo {author} {\bibfnamefont {X.}~\bibnamefont
  {Schlagberger}}, \bibinfo {author} {\bibfnamefont {Y.-W.}\ \bibnamefont
  {Kim}}, \ and\ \bibinfo {author} {\bibfnamefont {R.~R.}\ \bibnamefont
  {Netz}},\ }\href {\doibase 10.1039/b516777a} {\bibfield  {journal} {\bibinfo
  {journal} {Soft Matter}\ }\textbf {\bibinfo {volume} {2}},\ \bibinfo {pages}
  {653} (\bibinfo {year} {2006})}\BibitemShut {NoStop}%
\end{thebibliography}
\end{document}